\documentclass[a4paper,11pt]{article}
\pdfoutput=1 % 

\usepackage{jheppub} 
\usepackage{amsmath,amssymb,graphicx,float,slashed,xcolor,multicol}
\usepackage{tabularx}
\usepackage{url}
\usepackage{footmisc}
\usepackage{amsfonts}
\usepackage{cancel}
\usepackage{color}
\usepackage{multirow} 
\usepackage{pifont}
\usepackage{epstopdf}
\usepackage{comment}
\usepackage{booktabs} 
\usepackage{natbib}
\usepackage{array}
\usepackage{mathrsfs}
\usepackage[toc,page]{appendix}
\usepackage{mathtools}
\usepackage{romannum}
\usepackage[normalem]{ulem}
\usepackage{bbold}
\usepackage{enumitem}
\usepackage{multirow}
\usepackage{cleveref}
\usepackage{caption}
\usepackage{subcaption}
\usepackage{float,placeins}
\usepackage{multirow}
\usepackage{upgreek}
\usepackage[toc,page]{appendix}
\usepackage{romannum}
\allowdisplaybreaks
\usepackage{bm}
\usepackage{bbm}                % for \mathbbm{1} (unit matrix)
\usepackage{xspace}				% For spacing after command

\usepackage[compat=1.0.0]{tikz-feynman}
\usetikzlibrary{arrows,shapes}
\usetikzlibrary{trees}
\usetikzlibrary{matrix} 
\usetikzlibrary{positioning}				% For "above of=" commands
\usetikzlibrary{calc,through}				% For coordinates
\usetikzlibrary{decorations.pathreplacing}  % For curly braces
\usepackage{pgffor}							% For repeating patterns
\usetikzlibrary{decorations.pathmorphing}	% For Feynman Diagrams
\usetikzlibrary{decorations.markings}

\tikzstyle{block} = [draw, rectangle, 
minimum height=3em, minimum width=6em]

\newcommand*{\rom}[1]{\expandafter\@slowromancap\romannumeral #1@}

\newcommand{\cO}{\mathcal{O}}
\newcommand{\cL}{\mathcal{L}}
\newcommand{\cM}{\mathcal{M}}

\newcommand{\MeV}{\mathrm{MeV}}
\newcommand{\GeV}{\mathrm{GeV}}
\newcommand{\TeV}{\mathrm{TeV}}

\newcommand{\eg}{\textit{e.g.}}

\newcommand{\abs}[1]{| #1 |}

\def\lag{\mathscr{L}}

\def\beq{\begin{equation}}
\def\eeq{\end{equation}}
\def\beqa{\begin{eqnarray}}
\def\eeqa{\end{eqnarray}}

\newcommand{\Xtr}[1]{\left\langle #1 \right\rangle}

\title{Light scalar beyond the Higgs mixing limit}

\preprint{LAPTH-002/25,  CHIBA-EP-264}
\author[a]{C\'edric Delaunay,}
\emailAdd{cedric.delaunay@lapth.cnrs.fr}
\affiliation[a]{Laboratoire d'Annecy de Physique Th\'eorique, CNRS -- USMB, 74940 Annecy, France}

\author[b,c]{Teppei Kitahara,}
\emailAdd{kitahara@chiba-u.jp}
\affiliation[b]{Department of Physics, Graduate School of Science,
Chiba University, Chiba 263-8522, Japan}
\affiliation[c]{Kobayashi-Maskawa Institute for the Origin of Particles and the Universe, Nagoya University, Nagoya 464-8602, Japan}

\author[d]{Yotam Soreq}
\emailAdd{soreqy@physics.technion.ac.il}
\affiliation[d]{Physics Department, Technion -- Israel Institute of Technology, Haifa 3200003, Israel}

\author[e]{and Jure Zupan}
\emailAdd{zupanje@ucmail.uc.edu}
\affiliation[e]{Department of Physics, University of Cincinnati, Cincinnati, Ohio 45221,USA}
\abstract{
We explore the possibility that the interactions of a light scalar singlet, which mixes with the Standard Model~(SM) Higgs, also receive other UV contributions of comparable size. 
We  focus, in particular, on the flavor aligned limit, where couplings of the light scalar to the SM are almost flavor diagonal, but not necessarily proportional to the Higgs Yukawa couplings. 
The phenomenology of such a general flavor aligned light scalar differs from both the Higgs-mixed scalar, as well as from a general axion-like particle. 
We explore this for light scalar masses below a few hundred MeV, such that they can be produced in kaon decays, and in decays of $\eta$ and $\eta'$ mesons, and the transitions described using chiral perturbation theory. 
We then derive constraints on the light scalar interactions, assuming that light scalar decays are either just into photons or are invisible. 
We also discuss several UV examples of such light scalar models: 
a two-Higgs doublet model extended by a light scalar, a light dilaton from the dark sector, and a SM extended by heavy vector-like quarks and a light scalar. For the latter we also performed matching onto low energy theory at one-loop. 
}
\begin{document}
	
\titlepage
\maketitle

%%%%%%%%%%%%%%%%%%%%%%%%%%%%%%%%%%%%
\flushbottom
%%%%%%%%%%%%%%%%%%%%%%%%%%%%%%%%%%%%
\newpage
\begin{flushright}
\begin{minipage}{0.45\textwidth}
{\em 
Imagine there's a new physics heaven. \\
It is easy if you try a new light scalar, $\phi$. \\
John Lennon (paraphrased), c. 2025.
}
\end{minipage}
\end{flushright}
	
%%%%%%%%%%%%%%%%%%%%%%%%%%%%%%%%%%%%
\section{Introduction}
\label{sec:intro}
%%%%%%%%%%%%%%%%%%%%%%%%%%%%%%%%%%%%

A light spin-0 particle $\phi$ features in many well-motivated beyond the standard model~(BSM) scenarios. 
Its mass, $m_\phi$, is protected against large UV corrections, if $\phi$ is a pseudo Nambu-Goldstone boson~(pNGB) of a spontaneously broken global U(1) symmetry. 
In that case, $m_\phi$ can be well below the U(1) symmetry breaking scale $f$ without tuning (without a hierarchy problem), because the value of $m_\phi$ is controlled by the small explicit breaking of the U(1) symmetry, and not by the UV physics. 
The well known examples of such pNGBs include light CP-even scalars such as a dilaton mimicking the SM Higgs~\cite{Goldberger:2007zk} and a Randall-Sundrum radion~\cite{Randall:1999ee,Goldberger:1999uk,Goldberger:1999un}, as well as CP-odd scalars such as the strong CP problem axion and the axion-like-particles~(ALPs) (for reviews, see,~\eg~\cite{DiLuzio:2020wdo,Cirelli:2024ssz,OHare:2024nmr}), and a relaxion~\cite{Graham:2015cka,Choi:2016luu,Flacke:2016szy,Banerjee:2020kww}.

If one ignores the hierarchy problem, one can also contemplate light scalar scenarios in which the $\phi$ mass is not protected against the UV corrections. 
The most discussed BSM example of this type is a light scalar that mixes with the standard model~(SM) Higgs either through trilinear or quartic couplings, so that at low energies all the couplings to the SM fermions, gluons and photons are then fixed in terms of just one parameter, the mixing angle $\theta$~\cite{Patt:2006fw,Beacham:2019nyx}. 
In the light Higgs-mixed scalar limit the couplings between $\phi$ and the SM fermions are still proportional to the SM fermion masses (as they are for the Higgs), but with their strengths reduced by the  mixing angle $\theta$. 
However, this is certainly not the most general possibility.
Especially for the couplings of $\phi$ to the light SM fermions, $e$, $\mu$, $u$, $d$, $s$, which are suppressed by small Yukawa couplings $y_f\sim 10^{-5}-10^{-3}$, it is quite possible that higher dimensional operators could give larger contributions than the mixing with the Higgs.
Such corrections can then result in a drastically modified light scalar phenomenology. 

In this manuscript, we focus on the so called general flavor aligned limit, which assumes that the flavor off-diagonal couplings of $\phi$ to the SM fermions are small. 
The couplings of $\phi$ to the SM fermions are thus mostly diagonal, as in the light Higgs mixed scalar, however, the values of the diagonal Yukawa couplings can deviate from the Higgs-mixed scalar predictions, for previous studies in this direction, see \cite{Batell:2017kty,Batell:2018fqo,Batell:2021xsi,DiLuzio:2020oah,Balkin:2024qtf}. 
Importantly, the general flavor aligned scalar limit cannot be reached in the most general ALP effective Lagrangian, since the derivative ALP couplings result in vanishing flavor diagonal scalar (vector) couplings to SM fermions even in the most general case. 

One of the goals of the present work is to perform an initial study of the general flavor aligned light scalar phenomenology. 
We limit the discussion to $m_\phi$ masses to below a few 100\,MeV, which simplifies the analysis since it limits both the possible decay modes of $\phi$ as well as the production channels. 
We are especially interested in the regime where $K\to \pi \phi$ decays are kinematically allowed, since these decays are one of the most sensitive probes of the new sub-GeV particles. 
The other aim is to provide a set of UV BSM examples that contain a light general flavor aligned scalar.

The paper is organized as follows. 
In \cref{sec:BSMint} we first introduce the interactions of a general light scalar within an effective field theory framework, supplementing SM Effective Field Theory~(SMEFT) with a light scalar, and then discuss different limiting cases: 
the light Higgs-mixed scalar, the general ALP scenario, and the general flavor aligned light scalar case. 
In \cref{sec:ew:EFT} we discuss the low energy interactions of $\phi$, first within the weak effective theory, and then at the level of a chiral perturbation theory, with the aim to describe nonleptonic decays of kaons, $\eta$, and $\eta'$, involving $\phi$, \cref{sec:chiPT}. 
\Cref{sec:UV} contains examples of UV models for a general light scalar, while in \cref{sec:bounds} we translate the experimental bounds into constraints on the couplings of the light scalar.  
\Cref{sec:outro} contains our conclusions, while further details are relegated to a number of appendices; the details about the light neutral meson mass matrix diagonalization are given in \cref{sec:pietamixing}, \cref{sec:ChPT:derivation} gives details about the derivation of ChPT Lagrangian for $\phi$ interactions, \cref{sec:Feynman-rules} lists the Feynman rules for the vector-like quark (VLQ) model, while \cref{sec:renormalization} discusses   the penguin contributions in the case of down-like  VLQs.

%%%%%%%%%%%%%%%%%%%%%%%%%%%%%%%%%%%%
\section{A general light scalar}
\label{sec:BSMint}
%%%%%%%%%%%%%%%%%%%%%%%%%%%%%%%%%%%%

%%%%%%%%%%%%%%%%%%%%%%%%%%%%%%%%%%%%
\subsection{SMEFT${}_\phi$}
\label{sec:SMEFTphi}
%%%%%%%%%%%%%%%%%%%%%%%%%%%%%%%%%%%%

Let us first consider interactions of a general light scalar as described by the ``SMEFT${}_\phi$'' effective field theory, i.e., the SMEFT supplemented by a light scalar $\phi$, a gauge singlet under the SM. 
Any other degrees of freedom, apart from the SM fields and $\phi$, are assumed to be parametrically heavier than the electroweak~(EW) scale, $\mu_{\rm EW}\approx 100\,\GeV$, and have been integrated out. 
At $\mu=\mu_{\rm EW}$, the SMEFT${}_\phi$ Lagrangian is given by,
\begin{align}
    \label{eq:LSMEFTphi}
    \cL_{{\rm SMEFT}_\phi}
    = 
    \cL_{\rm SMEFT}
    +\frac{1}{2}(\partial_\mu\phi)^2-\frac{1}{2}m_\phi^2\phi^2
    +\frac{\phi}{f} \sum_i \eta_i \cO_i+\cdots\,,
\end{align}
where the form of the SMEFT Lagrangian, $\cL_{\rm SMEFT}$, can be found, e.g., in Ref.~\cite{Grzadkowski:2010es} for operators up to dimension-six. 
The sum in \cref{eq:LSMEFTphi} runs over all the higher-dimension operators linear in $\phi$, where $\cO_i$ are the scalar currents constructed out of the SM fields (derivable from results in Ref.~\cite{Song:2023lxf}). 
The Wilson coefficients $\eta_i$ multiplying the operators are dimensionless for dimension-four SM operators $\cO_i$, and are dimensionful otherwise. 
The ellipses in \cref{eq:LSMEFTphi} denote operators at least quadratic in $\phi$, which are not relevant to our analysis.  

For clarity, we distinguish two scales in SMEFT${}_\phi$: 
$f$ is the UV scale associated with the dynamics of $\phi$, while $\Lambda$ is a common mass scale of any other heavy fields (the two can also coincide). 
For instance, for $\phi$ that is a pNGB, $f$ can be identified with the scale at which the global symmetry is spontaneously broken. 
For reasons that will become apparent below, we work in a basis where the operators with a derivative acting on $\phi$ have been removed by performing integration by parts.  
Furthermore, as is usual in SMEFT, the Higgs doublet is assumed to be fully responsible for the EW symmetry breaking, and is thus in the unitary gauge given by $H=(0, v+h)/\sqrt2$, where $v\simeq 246\,\GeV$ is the Higgs VEV, and $h$ the would-be SM Higgs boson. Note that, while this assumption agrees well with data, it can be relaxed if needed.  In that case, for instance, one would replace SMEFT${}_\phi$ with HEFT${}_\phi$, the Higgs Effective Field Theory~(HEFT), supplemented by a light scalar $\phi$.

Next, let us explore what the different couplings in SMEFT${}_\phi$ imply for the phenomenology of $\phi$. 
At the renormalizable level, $\phi$ couples to the $|H|^2$ operator, which, after EW symmetry breaking, leads to mass mixing between the weak eigenstates $\phi$ and $h$~\cite{Patt:2006fw,Beacham:2019nyx}. 
Performing mass diagonalization gives,  
\begin{align}
    \label{eq:mass:mixing}
    \phi = \hat \phi \cos\theta - \hat h \sin\theta\, , 
    \qquad 
    h = \hat \phi \sin\theta  +\hat  h \cos\theta \, ,
\end{align}
where the mass eigenstates $\hat h$ and $\hat \phi$ can be identified with the observed Higgs-like state of mass $m_h\simeq125\,\GeV$ and the light scalar, respectively. 
The experimental constraints, \eg~\cite{CMS:2018uag,ATLAS:2019nkf}, require the mixing angle $\theta$ to be small, and we thus approximate $\cos\theta\simeq 1$. 

The interactions of $\hat{\phi}$ with the SM fermions and gauge fields thus come from \textit{two distinct sources}. 
Firstly, the $\hat{\phi}$ couplings to the SM fermions and gauge fields are generated from the Higgs Yukawa couplings and the Higgs couplings to the gauge bosons, via the $\phi-h$ mixing in \cref{eq:mass:mixing} (the $h$ couplings, in general, deviate from the SM Higgs couplings due to the corrections from the higher-dimension SMEFT operators). 
The second source of $\phi$ interactions with the SM fields are the genuine SMEFT${}_\phi$  operators of the form $\phi\, \cO_i$, which, in general, have no correlation with the Higgs interactions. 
This then combines to give a SMEFT${}_\phi$ Lagrangian after EW symmetry breaking at $\mu=\mu_{\rm EW}$
\begin{align}
    \label{eq:SMEFT:phi:full}
    \cL_{{\rm SMEFT}_\phi}^{\slashed{\rm EW}}
    =
    \cL_{{\rm SMEFT}_\phi}^{\rm gauge}
    +\cL_{{\rm SMEFT}_\phi}^{\rm ferm}\,,
\end{align}
for couplings of $\phi$ to SM gauge bosons and fermions, respectively.  

For instance, the leading contributions to the $\phi$ interactions with the SM gauge bosons that are induced by the Higgs-$\phi$ mixing, come from the Higgs kinetic term $|D_\mu H|^2$, and from the dimension six SMEFT operators $H^2F^2$ and $H^2F\tilde F$. 
The SMEFT${}_\phi$ contributions from the operators containing $\phi$ (``the direct SMEFT${}_\phi$ contributions''), start instead at dimension-five, and are due to the operators $\phi |D_\mu H|^2$, $\phi F^2$ and $\phi F\tilde F$. 
After EW symmetry breaking, the Lagrangian describing couplings of $\phi$ to gauge bosons is thus given by,\footnote{For gluons, a summation over color is understood, i.e., $V_{\mu\nu}V^{\mu\nu}\to G^a_{\mu\nu} G^{a\mu\nu}$, while for $W$ the coupling is between $W^+$ and $W^-$, i.e., $V_\mu V^\mu \to W_\mu^+ W^{-\mu}$, and we use the conventional $F_{\mu\nu}$ notation for the electromagnetic field strength.}
\beq
    \label{eq:LSMEFT:phi:gauge}  
    \begin{split}
    \cL_{{\rm SMEFT}_\phi}^{\rm gauge}
    =&
    \frac{\phi}{f}\! \sum_{V=G,\gamma,Z,W}   
    \!\!\!\!\!\!\!\Big(
    2\kappa_Vm_V^2 V_\mu V^\mu
    + c_V V_{\mu\nu}V^{\mu\nu} +\tilde c_V V_{\mu\nu}\tilde V^{\mu\nu} \Big)
    \\
    &+ \frac{\phi}{f} \Big(c_{Z\gamma}Z_{\mu\nu}F^{\mu\nu}+\tilde c_{Z\gamma} Z_{\mu\nu}\tilde F^{\mu\nu}\Big)\,,
    \end{split}  
\eeq
where $F_{\mu\nu}$ and $\tilde{F}_{\mu\nu}=\frac{1}{2}\epsilon_{\mu\nu\alpha\beta} F^{\alpha \beta}$ are the photon field strength and its dual, with similar definitions for gluons ($G^a$), as well as for $W$ and $Z$ bosons.  
To simplify the notation, we have also dropped the hat on $\hat \phi$, which is the notation we will use from now on (same for $\hat h$). 
That is, in the remainder of the manuscript $\phi$ and $h$ denote mass eigenstates, unless specified otherwise.
Each of the coefficients in \cref{eq:LSMEFT:phi:gauge} receives a Higgs-$\phi$ mixing contribution, 
\begin{align}
    \kappa_V
    =
    \frac{f}{v}\sin\theta+\cdots\,, 
    \qquad 
    c_V,\tilde c_V\sim \frac{vf}{\Lambda^2}\sin\theta+\cdots\,,
\end{align}
as well as a direct SMEFT${}_\phi$ contribution, here denoted with the ellipses. 
Note that for gluons and photons, $\kappa_{G}=\kappa_\gamma=0$. 

The interactions of $\phi$ with the SM fermions, being the main point of our analysis, require more scrutiny. 
The Higgs-mixing induced $\phi$ couplings to the SM fermions arise both from the renormalizable Yukawa couplings, as well as from the dimension-six SMEFT corrections to them (for the discussion of these, see, \eg, Ref.~\cite{Harnik:2012pb})
\beq
    \begin{split}
    \label{eq:SMEFT}
    \cL_{\rm SMEFT}
    &\supset 
    - \lambda^\psi_{ij} \bar \Psi_{Li} \psi_{Rj}H
    -\frac{\lambda_{ij}^{\psi\prime}}{\Lambda^2} \bar\Psi_{Li} \psi_{Rj}H\big(H^\dagger H)+{\rm h.c.}
    \\
    &\simeq
    - m_{\psi i} \bar \psi_{Li}\psi_{Ri}-\frac{Y^\psi_{ij}}{\sqrt{2}}\bar \psi_{Li} \psi_{Rj}(h + \phi\sin\theta)+{\rm h.c.}+\cdots.
    \end{split}
\eeq
Here, $\Psi_L$ and $\psi_R$ denote the SM SU(2)$_L$ doublets and singlets, respectively, and $\psi_L$ is the component of $\Psi_L$ that has the same electric charge as $\psi_R$. 
Implicit summations over $\psi=u,d,e$ and generation indices $i,j=1,2,3$ are understood. 
The Higgs Yukawa matrices are given by $Y\equiv V_L\big[\lambda+3 \lambda' v^2/(2 \Lambda^2)\big]V_R^\dagger$, where $V_{L,R}$ are the unitary matrices that diagonalize the fermion mass matrix (we dropped the superscript $\psi$ for brevity). 

The direct SMEFT${}_\phi$ contributions, on the other hand, arise from dimension-five couplings of the form $\phi \bar\psi\psi H$. 
In the mass basis, the interaction Lagrangian $\cL_{{\rm SMEFT}_\phi}^{\rm ferm}$ is therefore given by
\begin{align}
    \label{eq:SMEFTphi:ferm}
    \cL_{{\rm SMEFT}_\phi}^{\rm ferm}
    =& 
    - \frac{\phi}{f} m_{\psi_{ij}} \bar\psi_i\big(  \kappa_{ij}^\psi
    +i \tilde \kappa_{ij}^\psi\gamma_5\big) \psi_j, 
\end{align}
where  $m_{\psi_{ij}} \equiv (m_{\psi_i}m_{\psi_j})^{1/2} $ and
\begin{align}
    \label{eq:Yijphi}
    \kappa_{ij}^\psi
    &=
    \frac{1}{2\sqrt{2}}\frac{f}{m_{\psi_{ij}}}\biggr[\Big(Y^\psi_{ij}+Y^{\psi*}_{ji}\Big)\sin \theta +\Big(X^\psi_{ij}+X^{\psi*}_{ji}\Big)\frac{v}{f}\biggr]\,,
    \\
    \label{eq:tildeYijphi} 
    \tilde \kappa_{ij}^\psi 
    &= 
    -\frac{i}{2\sqrt{2}} \frac{f}{m_{\psi_{ij}}} 
    \biggr[\Big(Y^\psi_{ij}-Y^{\psi*}_{ji}\Big)\sin\theta+\Big(X^\psi_{ij}
    -X^{\psi*}_{ji}\Big)\frac{v}{f}\biggr]\, , 
\end{align}
with $X^\psi_{ij},\tilde X^\psi_{ij}$ the direct SMEFT${}_\phi$ contributions from the $\phi\,\cO_i$ terms. The $\kappa_{ij}^\psi$ and $\tilde \kappa_{ij}^\psi$ coefficients are hermitian $\kappa^\psi_{ji} = \kappa^{\psi*}_{ij}$, $\tilde\kappa^\psi_{ji} = \tilde\kappa^{\psi*}_{ij}$.
Note that the Higgs-mixing term, $\propto Y^\psi_{ij} \sin\theta$ in \cref{eq:Yijphi}, is proportional to the fermion masses, $Y^\psi_{ij}\propto m_{\psi_i} \delta_{ij}/v$, only in the $\Lambda\to \infty$ limit, while in general the $1/\Lambda^2$ contributions from the $\bar\psi_i\psi_j H^3$ SMEFT operators can lead to phenomenologically important corrections. 
Similarly, the flavor structure of the $X^\psi_{ij}$ couplings can be completely unrelated to the SM Yukawa couplings, and can in principle modify the  interactions of $\phi$ with the SM fermions far away from the minimal Higgs-mixed scalar limit. 

%%%%%%%%%%%%%%%%%%%%%%%%%%%%%%%%%%%%
\subsection{Limiting cases}
\label{sec:limiting:cases}
%%%%%%%%%%%%%%%%%%%%%%%%%%%%%%%%%%%%

From a purely EFT point of view there are many choices for the values of the $Y^\psi_{ij}$ and $X^\psi_{ij}$ couplings. 
It is thus instructive to consider three distinct limiting cases:
\begin{itemize}
\item 
\textbf{The light Higgs-mixed scalar limit} is obtained when the SM singlet $\phi$ interacts with the SM only via the two renormalizable couplings to the Higgs, the trilinear coupling $\phi |H|^2$ and the quartic coupling $\phi^2 |H|^2$~\cite{Patt:2006fw,Beacham:2019nyx}. 
In particular, one assumes that there are no other corrections to the Yukawa couplings (such as SMEFT higher dimensional operators) apart from these two mixing terms. 
After the EW symmetry breaking the two mixing terms generate $\phi$-Higgs mixing, which induces the following nonzero couplings of $\phi$ to the SM fermions in $\cL_{{\rm SMEFT}_\phi}^{\rm ferm}$ in \cref{eq:SMEFTphi:ferm}, and to the $V=Z, W$ gauge bosons in  $\cL_{{\rm SMEFT}_\phi}^{\rm gauge}$ in  \cref{eq:LSMEFT:phi:gauge}, giving 
\begin{align}
    \label{eq:Higgs-mixed}
    \kappa^\psi_{ii} = \frac{f}{v}\sin\theta \, , 
    \qquad 
    \kappa_{V} = \frac{f}{v}\sin\theta \, .
\end{align}
The flavor-violating couplings to the SM fermions, $\kappa_{i\ne j}^\psi$, as well as the CP-violating couplings, $\tilde \kappa_{ij}^\psi$, are both GIM and loop suppressed, and are thus negligible. 
At $\mu\sim \mu_{\rm EW}$ the couplings of $\phi$ to the photons and gluons in the SMEFT Lagrangian \cref{eq:SMEFT} are zero, $c_G=c_\gamma=0$ (and also trivially $\kappa_G=\kappa_\gamma=0$, because gluons and photons are massless).
At low energies the couplings to gluons and photons are generated when the heavy SM particles $c,t,b,W$ are integrated out, see \cref{eq:cG:phi,eq:cgamma:phi} in \cref{sec:EFT} below.
\item 
\textbf{The axion-like particle~(ALP) limit} assumes that $\phi$ is a pNGB of a spontaneously broken global symmetry, and thus has only derivative couplings to the SM fermions, $\partial_\mu \phi (\overline f_i \gamma^\mu f_j)$ and/or $\partial_\mu \phi (\overline f_i \gamma^\mu \gamma_5 f_j)$. 
The ALP couplings to gauge bosons are due to anomalies and are of the form $\phi G\tilde G$ and $\phi F\tilde F$ (see, \eg, Ref.~\cite{DiLuzio:2020wdo} and references therein).
This means that in the ALP limit, after using integration by parts, the scalar flavor-diagonal couplings to the SM fermions vanish, $\kappa_{ii}^\psi=0$ in \cref{eq:SMEFTphi:ferm}, as do the CP-conserving couplings to the gauge bosons, $c_{G,\gamma}=0$ in \cref{eq:LSMEFT:phi:gauge}. 
All the other couplings are in general nonzero, 
\begin{align}
    \kappa_{i\ne j}^\psi \ne 0 \, , \qquad 
    \tilde \kappa_{i j}^\psi \ne 0\, ,  \qquad 
    \tilde c_{G, \gamma}\ne 0\,,
\end{align}
though quite often the flavor off-diagonal couplings are also set to zero in the ALP literature, $\kappa_{i\ne j}^\psi =\tilde \kappa_{i \ne j}^\psi=0$ (see, \eg, Refs.~\cite{Irastorza:2018dyq,Choi:2020rgn,Ringwald:2014vqa}). 
Note, however, that flavor-violating ALP couplings, if nonzero, can be phenomenologically important~\cite{Wilczek:1982rv,Calibbi:2020jvd,Calibbi:2016hwq,Fuyuto:2024skf,Li:2024thq,Knapen:2023zgi,Hill:2023dym,Zhang:2023vva,DiLuzio:2023ndz,Ema:2016ops,Panci:2022wlc,Jho:2022snj,Carmona:2022jid,Goudzovski:2022vbt,DEramo:2021usm,Bauer:2021mvw,Cornella:2019uxs,Albrecht:2019zul,Bonnefoy:2019lsn,Bauer:2019gfk,Bjorkeroth:2018ipq,Bjorkeroth:2018dzu,Arias-Aragon:2017eww}. 
\item 
\textbf{The general flavor-aligned light scalar limit} is a scenario that we introduce in this manuscript. 
In this limit, the light scalar couplings are assumed to satisfy,
\begin{align}
    \label{eq:aligned:scalar}
    \kappa_{ii}^\psi \ne 0\,, \qquad  
    \kappa_{i\ne j}^\psi \ll \kappa_{ii}^\psi, \, \kappa_{jj}^\psi\,, \qquad
    c_{G,\gamma}\ne0\,,
\end{align}
with the diagonal couplings $\kappa_{ii}^\psi \ne 0$ having no specific structure, while the flavor off-diagonal couplings are assumed to be suppressed by approximate flavor symmetries in the UV.  
For simplicity, we also assumed that CP-odd couplings vanish, i.e., $\tilde \kappa^\psi_{ij}=\tilde c_{G,\gamma}=0$. 
\end{itemize}

Note that the general flavor-aligned light scalar limit is distinct from both the Higgs-mixed scalar limit (where all $\kappa_{ii}^\psi$ are controlled by a single free parameter), as well as from the ALP limit (in which the diagonal scalar couplings $\kappa_{ii}^\psi$ are zero). 
The minimality of the Higgs-mixed scalar carries a strong implicit assumption that the Higgs couplings to the first and second generations of fermions are the SM ones. 
This is not the case in many NP models, with several examples given in Sec.~\ref{sec:UV} below. 
In practical terms, because the experimental constraints on the Higgs Yukawa couplings to up and down quarks and electrons are rather weak~\cite{Kagan:2014ila,Bishara:2015cha,Perez:2015lra,Ghosh:2015gpa,Altmannshofer:2015qra,Chien:2015xha,Konig:2015qat,Soreq:2016rae,Bishara:2016jga,Altmannshofer:2016zrn,Egana-Ugrinovic:2019dqu,Brod:2022bww,Delaunay:2022grr}, the $\phi$ couplings to the first generation fermions can deviate significantly from the Higgs-mixed scalar pattern (this can occur even if the direct SMEFT${}_\phi$ couplings $X^\psi_{ij}$ in  \cref{eq:Yijphi,eq:tildeYijphi} vanish).
This can have drastic experimental consequences. 
For instance, unlike the light Higgs-mixed scalar it is no longer guaranteed that above the muon threshold the $\phi \to \mu^+\mu^-$ decays lead to the most stringent constraints, and thus one should also search for other signatures in this $\phi$ mass range.

In the rest of this manuscript, we develop the phenomenology of a general flavor-aligned light scalar, focusing on masses of up to a few hundred MeV so that kaon decays to $\phi$ are allowed, while leaving a full exploration of the phenomenology of a heavier $\phi$ for future work. 

%%%%%%%%%%%%%%%%%%%%%%%%%%%%%%%%%%%%
\section{Low energy interactions of $\phi$}
\label{sec:ew:EFT}
%%%%%%%%%%%%%%%%%%%%%%%%%%%%%%%%%%%%

For $\phi$ lighter than a few hundred MeV one can use Chiral Perturbation Theory~(ChPT) to describe interactions of $\phi$ with pions and kaons, as well as with $\eta^{(\prime)}$. 
To arrive at this ChPT description, we first write down the low energy Weak Effective Theory supplemented by $\phi$~(WET${}_\phi$), at the conventional renormalization scale of $\mu=\mu_{\rm had.} = 2\,\GeV$. 
This choice of scale is commonly used in lattice QCD calculations of the required hadronic matrix elements and low energy constants, which then facilitates the non-perturbative matching of WET${}_\phi$ on ChPT for $\phi$ interactions. 

%%%%%%%%%%%%%%%%%%%%%%%%%%%%%%%%%%%%
\subsection{WET${}_\phi$}
\label{sec:EFT}
%%%%%%%%%%%%%%%%%%%%%%%%%%%%%%%%%%%%

The WET${}_\phi$ effective theory is obtained from SMEFT${}_\phi$, \cref{eq:LSMEFTphi}, by integrating out the heavy degrees of freedom: $W$ and $Z$ gauge bosons, as well as the heavy quarks, $t, b$ and $c$, at the corresponding mass thresholds, and perform renormalization group (RG) running down to $\mu_{\rm had.}$. 
We will not require the full WET${}_\phi$ basis, but rather only a subset of operators,
\begin{align}
    \label{eq:WET:Lagr}
    \cL_{\text{WET}{}_\phi}
    =
    \cL^{\Delta S=1}_{4q}+\cL_{\rm int}^{\rm diag}+ \cL_{\rm int}^{sd}+\cdots,
\end{align}
where we discuss the different contributions relevant for our analysis in detail below. 

Note that the WET${}_\phi$ includes as part of it the WET basis, i.e., the operators that do not contain the $\phi$ field. 
For our analysis we only need the $\Delta S=1$ part of the WET Lagrangian, $\cL_{\text{WET}}\supset \cL^{\Delta S=1}_{4q}$, i.e., the four-quark operators that induce the $K\to \pi\pi$ decays.
Using the notation in Ref.~\cite{Buchalla:1995vs} these are given by
\beq
    \label{eq:DeltaS=1}
    \cL^{\Delta S=1}_{4q}
    =
    -\frac{G_F}{\sqrt 2} V_{us}^* V_{ud} 
    \sum_{i=1}^{10}  C_i(\mu) Q_i(\mu)\,,
\eeq
with
\beq
    Q_1=(\bar s_\alpha u_\beta)_{V-A} (\bar u^\beta d^\alpha)_{V-A}\,, 
    \qquad 
    Q_2=(\bar s u)_{V-A} (\bar u d)_{V-A}\,,
\label{eq:Q12}
\eeq
where $(V-A)$ refers to the $\gamma_\mu (1- \gamma_5)$ Dirac structure in the quark currents, and the summation over the QCD indices $\alpha, \beta$ is understood. 
The $Q_{1,2}$ four fermion operators are generated at tree level by integrating out the $W$ boson at $\mu=\mu_{\rm EW} \simeq m_W$. 
Performing the QCD RG evolution from $\mu_{\rm EW}$ down to $\mu_{\rm had.}$ gives $C_1(\mu_{\rm had.})= -0.309$, $C_2(\mu_{\rm had.})=1.145 $ at next-to-leading order~(NLO)~\cite{Buchalla:1995vs}. 
The forms of the QCD penguin operators, $Q_{3,\ldots,6}$, and the electroweak penguins, $Q_{7,\ldots,10}$, can be found in Ref.~\cite{Buchalla:1995vs}. 
While these are relevant when predicting the $K\to \pi\pi$ rates within the SM, they are not required for our purposes, and thus from now on we only keep the tree level operators, $Q_{1,2}$, in \cref{eq:DeltaS=1}.

From the full basis of WET${}_\phi$ operators containing the $\phi$ field, we also keep just a subset of operators, $\cL_{{\rm WET}_\phi}\supset \cL_{\rm int}^{\rm diag} + \cL_{\rm int}^{sd}$, see \cref{eq:WET:Lagr}.
For flavor diagonal couplings of $\phi$ to the SM fermions and gauge bosons, we keep only the operators of the lowest dimension 
\begin{align}
    \label{eq:Lint2GeV}
    \cL_{\rm int}^{\rm diag}
=   \frac{\phi}{f}\biggr(
    c_G^\phi G_{\mu\nu}^a G^{a\,\mu\nu}
    +c_\gamma^\phi F_{\mu\nu}F^{\mu\nu}
    -\sum_{\psi=q,\ell}\kappa_\psi^\phi  m_\psi\overline \psi \psi \biggr)\,,
\end{align}
where the sum is over the light quark flavors, $q=u,d,s$, and the two lightest charged leptons $\ell=e, \mu$.  We also limit the discussion to just the CP even interactions. 
From the complete set of $s\to d$ transition operator we keep, for reasons that will become apparent below, both the two fermion and the four fermion operators
\begin{align}
    \label{eq:Q12L}
    \cL_{\rm int}^{sd}
    =
    \frac{\phi}{f}\big(\kappa_{sd}^\phi m_s \bar d_L s_R 
    + \kappa_{ds}^\phi m_d \bar d_R s_L+{\rm h.c.}\big) 
    +\frac{\phi}{f}\sum_{i=1,2} c_i^\phi(\mu) Q_i(\mu)\,,
\end{align}
where $Q_i$ are given in \cref{eq:Q12}. Note that in general $\kappa_{sd}^\phi$ and $\kappa_{ds}^\phi$ are complex. 

The flavor diagonal WET${}_\phi$ couplings to light SM fermions in \cref{eq:Lint2GeV} are given in terms of the SMEFT${}_\phi$ Wilson coefficients in \cref{eq:SMEFTphi:ferm} as
\beq
    \label{eq:kappa:phi:q}
    \kappa_u^\phi=\kappa_{11}^u, \qquad 
    \kappa_d^\phi=\kappa_{11}^d, \qquad 
    \kappa_s^\phi=\kappa_{22}^d,\qquad 
    \text{and}\qquad
    \kappa_{e, \mu}^\phi=\kappa_{11,22}^\ell\, .
\eeq
The WET${}_\phi$ Wilson coefficients for $\phi$ couplings to gauge bosons in \cref{eq:Lint2GeV}, on the other hand, are 
\begin{align}
    \label{eq:cG:phi} 
    c_G^\phi
    =& 
    c_G+\frac{\alpha_s}{12\pi}\sum_Q \kappa_Q^\phi \,, \\
    \label{eq:cgamma:phi}
    c_\gamma^\phi
    =& 
    c_\gamma+\frac{\alpha}{6\pi}\biggr(3\sum_Q\kappa_Q^\phi Q_Q^2+\kappa_\tau^\phi-\frac{21}{4}\kappa_W+c_W r_W\biggr)\,, 
\end{align}
where $c_{G,\gamma,W}$, $\kappa_W$ are the SMEFT${}_\phi$ Wilson coefficients for couplings to gauge bosons, \cref{eq:LSMEFT:phi:gauge}.
The summation in \cref{eq:cG:phi,eq:cgamma:phi} is over the heavy quarks that were integrated out, $Q=t,b,c$, with $Q_Q=2/3,-1/3,2/3$ the corresponding electric charges. Furthermore,  we shortened the expressions for summations using the same notation as in \cref{eq:kappa:phi:q}, 
\beq
    \label{eq:kappa:ctphi:WET}
    \kappa_{c,t}^\phi=\kappa_{22,33}^u, \qquad 
    \kappa_{b}^\phi=\kappa_{33}^d, \qquad 
    \kappa_{\tau}^\phi=\kappa_{33}^\ell. 
\eeq
The  $r_W$ in \cref{eq:cgamma:phi} is a regularization scheme dependent dimensionless constant, where typically $r_W\sim\cO(1)$ .

For $\cL_{\rm int}^{sd}$ part of the WET${}_\phi$ Lagrangian, the $c_i^\phi(\mu)$ Wilson coefficients in \cref{eq:Q12L} arise from tree-level $W$ exchange. Working in the soft $\phi$ limit, the $\kappa_W$ interaction in \cref{eq:LSMEFT:phi:gauge}  corrects the $W$ propagator as $1/[m_W^2(1+2\kappa_W\phi/f)]\simeq (1-2\kappa_W\phi/f)/m_W^{2}$. 
This then gives 
\begin{align}
    \label{eq:ciphi}
    c_i^\phi(\mu)
    =
    \sqrt{2} G_F V_{us}^* V_{ud} \kappa_W C_i(\mu)+\cdots,\qquad i=1,2,
\end{align}
where ellipses denote the other SMEFT${}_\phi$ corrections, which are expected to be suppressed by $m_W^2/f^2$. 
Keeping only the tree-level four-fermion operators $Q_{1,2}$ in the weak Lagrangian, we can thus write the corresponding terms in the $\phi$ interaction Lagrangian in \cref{eq:Q12L} as, 
\beq
    \cL_{\rm int}^{sd}
    \supset 
    \frac{\phi}{f}\sum_{i=1,2} c_i^\phi(\mu) Q_i(\mu)
    =
    -2 \kappa_W  \frac{\phi}{f} \cL^{\Delta S=1}_{4q}\, .
\eeq

The $sd\-\phi$ flavor-changing couplings in $\cL_{\rm int}^{sd}$ in \cref{eq:Q12L} are, on the other hand, given by (see also similar discussion for the case of an ALP in \cite{Gavela:2019wzg,MartinCamalich:2020dfe,Bauer:2020jbp,Chala:2020wvs})
\begin{align}
    \kappa_{sd}^\phi=\kappa_{sd}^{\text{cnt}}+\kappa_{sd}^{\rm loop}\,, 
    \qquad 
    \kappa_{ds}^\phi=\kappa_{ds}^{\text{cnt}}+\kappa_{ds}^{\rm loop},
\end{align}
where we perform a one-loop matching from SMEFT${}_\phi$ onto WET${}_\phi$.
The counter-terms $\kappa_{sd,ds}^{\rm cnt}$ are due to the SMEFT${}_\phi$ point interactions $\kappa^d$ and $\tilde{\kappa}^d$ in \cref{eq:SMEFTphi:ferm}, 
\begin{align}
    \label{eq:kappasdds}
    \kappa_{sd}^{\text{cnt}}
    =-\sqrt{\frac{m_d}{m_s}}\big(\kappa_{12}^d+i \tilde \kappa_{12}^d\big)\,, 
    \qquad 
    \kappa_{ds}^{\text{cnt}}
    =-\sqrt{\frac{m_s}{m_d}}\big(\kappa_{12}^d-i \tilde \kappa_{12}^d\big)\,,
\end{align}
and thus implicitly depend on the heavy states that were integrated out when the UV model was matched onto SMEFT${}_\phi$. 
The $\kappa_{sd,ds}^{\rm loop}$, on the other hand, are due to one-loop SMEFT${}_\phi$ scalar-penguin diagrams such as, \eg, \cref{fig:penguin-up}\,(a) and (b), giving,
\beq
\begin{split}
    \label{eq:zeta}
  \kappa_{ds(sd)}^{\rm loop}
=   \frac{3G_Fm_W^2}{8\sqrt{2}\pi^2}
     \sum_{i=u,c,t} x_{i} V_{is}V_{id}^\ast \biggr[& \kappa_if_i(x_i)
     +\sum_{k=d,s} \kappa_k \big(f_k(x_i) \pm \delta f_k(x_i)\big)
     \\
     &+\kappa_W  f_W(x_i)
    -3c_Wf_W'(x_{i})\biggr]\,,
\end{split}    
\eeq
where $x_{i}\equiv m_i^2/m_W^2$, and in the numerics we take $x_u=0$.
The one-loop functions $f_{u,c,t}(x_i)$, $f_{d,s}(x_i)\pm \delta f_{d,s}(x_i)$, $f_W(x_i)$ and  $f_W'(x_i)$ arise from diagrams where the $\phi$ scalar gets emitted from the internal up-type quark line, from the external down and strange quark legs, and from the internal $W$ line, respectively. 
Since $m_\phi\ll m_W$ and $m_{u,d,s}\ll m_W$ we neglect any dependence on $m_\phi$ and $m_{u,d,s}$. 
The $x_i$ pre-factors in \cref{eq:zeta} reflects the suppression from the GIM mechanism. 
The $f_W'$ function is UV finite 
\begin{align}
    f_W'(x) &=  \frac{x^2-2x\log x-1}{6(x-1)^3}\,.
\end{align} 
The $f_{u,c,t}(x_i)$, $f_{d,s}(x_i)\pm \delta f_{d,s}(x_i) $, and $f_W(x_i)$ functions, on the other hand, are UV log-divergent. 
Only the combinations  $\kappa_{sd}^{\rm cnt}+\kappa_{sd}^{\rm loop}$ and $\kappa_{ds}^{\rm cnt}+\kappa_{ds}^{\rm loop}$ are finite, in general, though $\kappa_{sd}^{\rm loop}$ may still be finite in a particular model. 
For instance, the loop functions satisfy~\cite{Willey:1982mc,Grzadkowski:1983yp}  
\begin{align}
   \sum_{j=t,W} f_j(x_t)+ \sum_{k=d,s} \big(f_k(x_t)\pm \delta f_k(x_t)\big)=1\,,
\end{align}
and the same for $t\to c$. 
Since for the Higgs-mixed scalar $\kappa_t=\kappa_c=\kappa_s=\kappa_d=\kappa_W$ and all off-diagonal fermionic couplings vanish, this means that in this case $\kappa_{ds}^{\text{cnt}}=\kappa_{sd}^{\text{cnt}}=0$.
Another example, the SM extended with heavy vector-like quarks~(VLQ), for  which $\kappa_{ds}^{\text{cnt}},\kappa_{sd}^{\text{cnt}}\ne0$, is presented in \cref{sec:UV}. 

Note that the penguin contributions with the $u$-quark running in the loop in the full UV theory get split into two types of contributions in WET${}_\phi$. 
The first is a penguin-like contribution from contracting into a loop the $u$-quark legs in the $\phi \,Q_i$ operators in \cref{eq:Q12L} (the ``long-distance contribution''). 
The counter-terms $\kappa_{sd,ds}^{\text{cnt}}$, \cref{eq:SMEFTphi:ferm}, contain the mismatch between the UV divergent contributions from such $u$-quark loops in WET${}_\phi$ and in the full theory (the ``short-distance contribution''). 
Note that the above terminology of long-distance vs. short-distance is not very precise, and becomes even more problematic once the above WET${}_\phi$ Lagrangian gets matched onto ChPT, \cref{sec:Lchiral}.

Before we move on to the discussion of the chiral Lagrangian, it is useful to write out explicitly the values of WET${}_\phi$ Wilson coefficients, \cref{eq:Lint2GeV,eq:Q12L}, for the example of a Higgs-mixed scalar.  
The nonzero SMEFT${}_\phi$ Wilson coefficients for the case of a Higgs-mixed scalar are given in \cref{eq:Higgs-mixed}. 
Using these expressions in \cref{eq:cG:phi,eq:cgamma:phi}, along with $c_G=c_W=c_\gamma=0$, gives the following WET${}_\phi$ Wilson coefficients at $\mu=2$\,GeV in $\cL_{\rm int}^{\rm diag}$, \cref{eq:Lint2GeV}, for the Higgs-mixed scalar case (treating charm quark as heavy)
\begin{align}
    \label{eq:cG:phi:Higgsmixed}
    c_G^\phi
    =& 
    \frac{\alpha_s}{4\pi}\frac{f}{v} \sin\theta\,, \\
    \label{eq:cgamma:phi:Higgsmixed}
    c_\gamma^\phi
    =& 
   -\frac{5\alpha}{24\pi}\frac{f}{v}\sin\theta\,, \\
   \label{eq:kappapsi:phi:Higgsmixed}
   \kappa_\psi^\phi
   =
   &\frac{f}{v}\sin\theta, \qquad \psi=u,d,s,e,\mu,
\end{align}
while the Wilson coefficients in $\cL_{\rm int}^{sd}$, \cref{eq:Q12L}, are given by,  
\begin{align}
    \label{eq:kappasd:Higgsmixed}
    \kappa_{sd}^\phi
    =&
    \kappa_{ds}^\phi
    =   
    \frac{3G_F}{8\sqrt{2}\pi^2}
    \Big(m_t^2 V_{ts}V_{td}^\ast+m_c^2 V_{cs}V_{cd}^\ast \Big)
    \frac{f}{v}\sin\theta\,,  \\
    c_i^\phi(\mu)
    =&
    \sqrt{2} G_F V_{us}^* V_{ud}  C_i(\mu)\frac{f}{v}\sin\theta,\qquad i=1,2.
\end{align}
Note that the factor $f/v$ in the Wilson coefficients combines with the factor $1/f$ in the definitions of interaction Lagrangians $\cL_{\rm int}^{\rm diag}$ and $  \cL_{\rm int}^{sd}$, to give an overall suppression of $1/v$ (one could have, equivalently, simply chosen $f=v$ as the conventional suppression scale, which is an appropriate choice for this case).
 
%%%%%%%%%%%%%%%%%%%%%%%%%%%%%%%%%%%%
\subsection{Chiral Lagrangian}
\label{sec:Lchiral}
%%%%%%%%%%%%%%%%%%%%%%%%%%%%%%%%%%%%

Since we are interested in processes below the kaon mass, the interactions in $\cL_{\text{WET}{}_\phi}$, \cref{eq:WET:Lagr}, can be matched nonperturbatively onto a chiral Lagrangian
\beq
    \cL_{\rm eff}^\phi(U)
    =
   \cL_{\rm eff}^{\rm str.}(U)+\cL_{\rm eff}^{\Delta S=1}(U)+\cL_{\rm int}^\phi(U) \, ,
\eeq
where $\cL_{\rm int}^\phi(U)$ contains ChPT interactions involving $\phi$, while the first and the second term describe the strong interactions and the weak $\Delta S=1$ transitions in the SM Chiral Lagrangian~\cite{Gasser:1983ky,Gasser:1984gg}, respectively. 

The structure of $\cL_{\rm eff}(U)$ encodes the spontaneous breaking of  global flavor symmetry in the light-quark sector of QCD, $SU(3)_L\times SU(3)_R\to SU(3)_V$. 
The $U$ matrix of pNGBs,
\begin{align}
    U(x)
    =
    \exp\biggr(\frac{i\sqrt{2}}{f_\pi}\Pi(x)\biggr)\,,
\end{align}
transforms under $SU(3)_L\times SU(3)_R$ as $U\to LU R^\dagger$. 
Here, $\Pi = \lambda^a \pi^a + \sqrt{\tfrac{2}{3}} \eta_0$, summing over the Gell-Mann matrices $\lambda_a$ of the spontaneously broken $SU(3)$, $f_\pi\simeq 130\,$MeV is the pion decay constant~\cite{FlavourLatticeAveragingGroupFLAG:2021npn}, and for convenience we also include in $\Pi$ the $SU(3)$ singlet $\eta_0$ meson associated with the anomalous $U(1)_A$ symmetry. 
Explicitly, 
\begin{align}
    \label{eq:Pi:explicit}
    \Pi 
    =
    \begin{pmatrix}
    \pi^0+\frac{1}{\sqrt{3}}\eta_8+\sqrt{\frac{2}{3}}\eta_0 & 
    \sqrt{2}\pi^+ & 
    \sqrt{2}K^+ \\ 
    \sqrt{2}\pi^- & 
    -\pi^0+\frac{1}{\sqrt{3}}\eta_8+\sqrt{\frac{2}{3}}\eta_0 & 
    \sqrt{2}K^0 \\ 
    \sqrt{2}K^- & 
    \sqrt{2}\,\overline{K}^0 & 
    -\frac{2}{\sqrt{3}}\eta_8+\sqrt{\frac{2}{3}}\eta_0
    \end{pmatrix}\,.
\end{align}
Below, we first review the forms of $\cL_{\rm eff}^{\rm str.}(U)$ and $\cL_{\rm eff}^{\Delta S=1}(U)$, and then extend the analysis by including the $\phi$ interactions.

%%%%%%%%%%%%%%%%%%%%%%%%%%%%%%%%%%%%
\subsubsection{Strong and weak interactions}
\label{sec:StrongWeakInt}
%%%%%%%%%%%%%%%%%%%%%%%%%%%%%%%%%%%%

The QCD Lagrangian, $\cL_{\rm QCD}=\bar q i\slashed D q- \bar q_L M_q q_R +{\rm h.c.}$, where $q=(u,d,s)$, is formally invariant under the $SU(3)_L\times SU(3)_R$  transformations, if $M_q\equiv\text{diag}(m_u, m_d, m_s)$ is promoted to a spurion transforming as $M_q \to L M_q R^\dagger$. 
The  ChPT Lagrangian, constructed out of $U(x)$ fields, is thus also invariant under the $SU(3)_L\times SU(3)_R$ transformations, if $M_q$  is treated as a spurion. 
At leading order in the chiral expansion, i.e., at $\cO(p^2)$, the ChPT Lagrangian is given by
\begin{align}
    \label{eq:Leff}
    \cL_{\rm eff}^{\rm str.}(U)
    =   
    \frac{f_\pi^2}{8}\left[\big\langle{D_\mu U^\dagger D^\mu U}\big\rangle
    +\big\langle{\chi^\dagger U + U^\dagger\chi}\big\rangle\right]
    -V(\eta_0)\,,
\end{align}
where 
\beq
    \label{eq:Veta0}
    V(\eta_0)
    = 
    \frac{\mu_{\eta_0}^2}{2}\eta_0^2+\cdots \, ,
\eeq
and $\Xtr{\cdots}$ denotes the trace over SU(3) flavor indices. 
The U(1)$_{\rm em}$ covariant derivative is $D_\mu U\equiv \partial_\mu U + i e  A_\mu[Q_q, U]$, with $Q_q \equiv \text{diag}(2/3,-1/3,-1/3)$, and 
\beq
    \chi = 2 B_0 \, \text{diag}(m_u, m_d, m_s) = 2 B_0 M_q\,,
\eeq
with $B_0$ the bag parameter related to the QCD confinement scale. 
At $\cO(p^2)$ in chiral counting, the pion and kaon masses are given by $m_{\pi^0}^2=m_{\pi^\pm}^2= B_0 (m_u+m_d)$, $m_{K^0}^2=m_{\bar{K}^0}^2=B_0 (m_d+m_s)$, $m_{K^\pm}^2=B_0 (m_u+m_s)$. 
Note that the $\eta_0$ interactions are not constrained by spontaneous symmetry breaking, and thus the form of $V(\eta_0)$ is arbitrary, apart from being parity even, $V(\eta_0)=V(-\eta_0)$. 
In our analysis we ignore terms beyond the quadratic mass term in \cref{eq:Veta0}. 

Similarly, one can use the symmetry properties of the weak Lagrangian $\cL^{\Delta S=1}_{4q}$ in \cref{eq:DeltaS=1} to write down the corresponding ChPT Lagrangian for $\Delta S=1$ weak interactions. 
At $\cO(p^2)$ the dominant $\Delta I = 1/2$ part  is described by two terms~\cite{Leutwyler:1989xj,Kambor:1989tz}, 
\beq
    \label{eq:Leffweak}
    \cL_{\rm eff}^{\Delta S=1}(U)
    = 
    -\frac{f_\pi^2}{8}\big[
    \gamma_1\big\langle{\lambda_6\,D_\mu U^\dagger D^\mu U}\big\rangle
    +\gamma_2 \big\langle{\lambda_6 (\chi^\dagger U + U^\dagger \chi)}\big\rangle\big]\,,
\eeq
where $\lambda_6$ is the SU(3) Gell-Mann matrix projecting onto the $d,s$ subspace.
Neglecting CP violation, $\gamma_1$ and $\gamma_2$ are real constants, where experimentally, $|\gamma_1|\simeq 3.1\times 10^{-7}$ from $K_S \to \pi^+\pi^-$ and $K_S \to 2\pi^0$ decays~\cite{Leutwyler:1989xj}.\footnote{In terms of another commonly used notation for weak interactions,  $\gamma_1 = -2 f_\pi^2 G_8$~\cite{Cirigliano:2011ny}.}
The $\gamma_2$ constant, on the other hand, is not directly observable since it merely renormalizes the quark mass matrix in \cref{eq:Leff}, though this will no longer be true once $\phi$ interactions are included, see below. 
The expectation is that $|\gamma_2/\gamma_1|\ll 1$~\cite{Leutwyler:1989xj}.
The  interactions in $\cL_{\rm eff}^{\Delta S=1}$ proportional to $\gamma_1$ and $\gamma_2$ induce off-diagonal kinetic and mass terms, respectively. 

%%%%%%%%%%%%%%%%%%%%%%%%%%%%%%%%%%%%
\subsubsection{Scalar $\phi$ interactions}
\label{sec:CPevenInt}
%%%%%%%%%%%%%%%%%%%%%%%%%%%%%%%%%%%%

The ChPT Lagrangian for the interactions of a CP even scalar $\phi$ can be derived treating $\phi$ as an external field, and working to linear order in $\phi$. 
At LO in chiral expansion, the Lagrangian describing $\phi$ interactions with light mesons, light charged leptons, and photons, is given by (see \cref{sec:ChPT:derivation} for the derivation) 
\beq
\begin{split}
    \label{eq:Leffphi}
    \cL_{\rm int}^\phi(U)
    =
    &\frac{\phi}{f} \frac{f_\pi^2}{4}  \Big[ 
    \big\langle Z_D D_\mu U^\dagger D^\mu U\big\rangle
    +\big\langle\chi_\phi^\dagger U+ U^\dagger \chi_\phi\big\rangle\Big]-Z_{\eta_0} \frac{\phi}{f} V(\eta_0)
    \\
    &+\frac{\phi}{f} \Big[Z_\gamma F_{\mu\nu}F^{\mu\nu}- \sum_{\ell}\kappa_\ell^\phi m_\ell\bar \ell \ell \Big]\,,
    \end{split}
\eeq
where 
\begin{align}
    Z_D
    &= K_\Theta+\big(\kappa_W- 2K_\Theta\big)\gamma_1 \lambda_6\, , \\
    \chi_\phi
    &=
    4 B_0 \left(K_\Theta M_q + \frac{1}{4}M_\kappa 
    +\frac{1}{2}\left(\kappa_W-3K_\Theta\right)\gamma_2 M_q \lambda_6
    - \frac{1}{4}\gamma_2 M_\kappa \lambda_6\right)\,, \\
    Z_\gamma
    &= 
    c_\gamma^\phi + \frac{\alpha}{4 \pi} K_\Theta= c_\gamma^\phi+\frac{2}{9}\frac{\alpha}{\alpha_s}c_G^\phi \, , \\
    Z_{\eta_0}
    &=
    4 K_\Theta \, . 
\end{align} 
Above, we utilized the short-hand notation 
\beq
    \label{eq:KTheta:main:text}
    K_\Theta = \frac{8\pi}{ 9 \alpha_s }c_G^\phi\, ,
\eeq
that is also used in \cref{sec:ChPT:derivation}, while $M_\kappa$ is the following $3\times 3$ matrix  
\begin{align}
    \label{eq:cMK:longer}
    M_\kappa
    =
    \begin{pmatrix}
        (\kappa_u^\phi-K_\Theta) m_u & 0 & 0  \\
        0 & (\kappa_d^\phi-K_\Theta) m_d &-\kappa_{ds}^\phi m_d \\
        0& -\kappa_{sd}^{\phi *} m_s & (\kappa_s^\phi-K_\Theta)m_s
    \end{pmatrix} \, .
\end{align}
To shorten the notation we also expressed the $c_{1,2}^\phi(\mu)$ WET${}_\phi$ Wilson coefficients, \cref{eq:ciphi}, in terms of the common  SMEFT${}_\phi$ Wilson coefficient, $\kappa_W$. 
Note that the Lagrangian in \cref{eq:Leffphi} is given in a basis that is not yet the mass basis; the Lagrangian contains both kinetic and mass mixings, and thus one still needs to perform the diagonalization.  

Two further comments are in order.  
First, since $\langle \chi_\phi\rangle\ne 0$ the effective Lagrangian in \cref{eq:Leffphi} includes a tadpole for $\phi$, leading to a non-zero vacuum expectation value $\langle 0| \phi|0\rangle\propto f_\pi^2 \langle \chi_\phi +\chi_\phi^\dagger\rangle/f m_\phi^2$. 
This has no observable effect on the physics we are interested in: performing the field redefinition $\phi \to\phi^\prime =  \phi - \langle 0| \phi|0\rangle$ still leads to interaction Lagrangian for $\phi^\prime$ in \cref{eq:Leffphi}, while the vev of $\phi$ simply rescales the values of low energy constants in the SM ChPT Lagrangian.

Second, the effective Lagrangian includes both $K^0$-$\phi$ and $\bar K^0$-$\phi$ mixing interactions,
\beq
    \cL_{\rm int}^\phi(U)
    \supset 
    -\frac{i f_\pi}{2 f} \big[(\chi_\phi)_{sd}-(\chi_\phi)_{ds}^*\big]\phi K^0+{\rm h.c.},
\eeq
so that the $K^0$-$\bar K^0$ mixing amplitude, $M_{12}$, receives an additional contribution from two insertions of $\cL_{\rm int}^\phi(U)$, giving
$M_{12}=M_{12}^{\rm SM}+M_{12}^{\phi}$.  
In the limit of light $\phi$, $m_\phi \ll m_{K^0}$,  the new physics contribution is given by
\beq
    M_{12}^{\phi} 
    = 
    \frac{\langle K^0 |\mathcal{H}^{\Delta S =2}_\phi| \overline{K}^0 \rangle}
    {2 m_{K^0}}
    =  
    \frac{1}{8 m_K^3}\left\{\frac{f_\pi}{f}\big[(\chi_\phi)_{sd}^*-(\chi_\phi)_{ds}\big]\right\}^2\, .
\eeq
The contributions of $\phi$ to the kaon mass difference, $\Delta m_K\equiv m_{K_L}-m_{K_S}$, and to the indirect CP-violation parameter, $\epsilon_K$, are $\Delta m_K^\phi =2 \text{Re}\,M_{12}^{\phi}$ and $\epsilon_K^\phi\simeq \textrm{Im}\,M_{12}^{\phi}/(\sqrt{2} \Delta m_K^{\rm exp})$, respectively~\cite{ParticleDataGroup:2024cfk}. 
The bounds from $\Delta m_K$ and $\epsilon_K$ are given by 
\beq
    \left| \frac{\text{Re} \left[(\chi_\phi)_{sd} 
    - (\chi_\phi)_{ds}\right] }{f} \right|  \lesssim 3.2 \times 10^{-4}\,\MeV\,, 
\eeq
and
\beq
    -6.2 \times 10^{-11} \,\MeV^2 < 
    \frac{\text{Re} \left[(\chi_\phi)_{sd} - (\chi_\phi)_{ds}\right] }{f} 
    \frac{\text{Im} \left[(\chi_\phi)_{sd} + (\chi_\phi)_{ds}\right] }{f} 
    < 4.2 \times 10^{-11}\,\MeV^2\,,
\eeq
respectively. 
In deriving the above bounds we imposed
$|\Delta m_K^\phi| < \Delta m_K^{\rm exp}$, due to the sizable theoretical uncertainties from the charm-quark box diagrams and the long-distance contributions \cite{Aebischer:2022fld}, as well as  $-0.29 \times 10^{-3}  <  \epsilon_K^\phi < 0.43 \times 10^{-3}$~\cite{Brod:2019rzc}.
As we show in \cref{sec:bounds}, these bounds are weaker
than the NA62 bound from the searches for $K^+ \to \pi^+ \phi$ decays.

%%%%%%%%%%%%%%%%%%%%%%%%%%%%%%%%%%%%
\subsubsection{\texorpdfstring{Corrections at $\mathcal{O}(p^4)$}{O(p4)}}
%%%%%%%%%%%%%%%%%%%%%%%%%%%%%%%%%%%%

At $\cO(p^2)$ in ChPT counting the $K^+\to \pi^+ \phi$ decay amplitude depends on the coupling combinations $\kappa_d^\phi-\kappa_s^\phi$ and $\kappa_u^\phi-(\kappa_d^\phi+\kappa_s^\phi)/2$, while the $K_L\to \pi^0\phi$ decay amplitude depends on $\kappa_d^\phi-\kappa_s^\phi$, but not on $\kappa_u^\phi$ and $\kappa_d^\phi+\kappa_s^\phi$, see \cref{sec:KaonDecays}. 
For the case $\kappa_u^\phi=\kappa_d^\phi=\kappa_s^\phi$ ($\kappa_d^\phi=\kappa_s^\phi$) the $K^+\to \pi^+\phi$ ($K_L\to \pi^0\phi$) amplitude depends on the light-quark couplings only starting at $\cO(p^4)$. While these limiting cases are only a small part of the complete available parameter space for $\kappa_q^\phi$ values, they can  be motivated in particular UV models. 
Having  an estimate of the bounds on $\kappa_q^\phi$ values also for the limiting cases, can thus be of phenomenological interest. 

A complete discussion of $\cO(p^4)$ corrections is rather complex; there are 38 independent $\Delta S=1$ and $\Delta I=1/2$ operators at $\cO(p^4)$ in the weak ChPT Lagrangian~\cite{Kambor:1989tz}. 
A full study of $\cO(p^4)$ corrections would, furthermore, require one-loop diagrams constructed from $\cO(p^2)$ interactions, which is beyond the scope of our current analysis (for the case of an ALP, though, see \cite{Cornella:2023kjq}). 

To obtain rough guidance on the $K^+\to\pi^+\phi$ decay amplitude's sensitivity to the light-quark couplings in the $\kappa_u^\phi=\kappa_d^\phi=\kappa_s^\phi$ limit, and on the $K_L\to \pi^0 \phi$ decay amplitude's sensitivity to $\kappa_u^\phi$ and the $\kappa_d^\phi+\kappa_s^\phi$ combination of light quark couplings, we calculate the transition probabilities induced by a single representative $\cO(p^4)$ operator, and leave a more complete analysis for future work. 
As the representative $\cO(p^4)$ operator we take,
\beq
    \cO_{14}^8
    =
    \langle \lambda_6 D_\mu U^\dagger  D^\mu U\rangle 
    \langle \chi^\dagger U +U^\dagger \chi\rangle\,,
\eeq
which results in contributions to the $K^+\to \pi^+ \phi$ and $K\to \pi^0 \phi$ decay amplitudes that depend on both $\kappa_u^\phi$ and $\kappa_d^\phi+\kappa_s^\phi$. 

Adding $\cO_{14}^8$ to \cref{eq:Leffweak}, gives the modified weak ChPT Lagrangian, 
\beq
    \lag_{\rm eff}^{\Delta S=1}(U)
    \to 
    \lag_{\rm eff}^{\Delta S=1}(U)+E_{14} \mathcal{O}_{14}^8\,.  
\eeq
Repeating the derivation of the $\phi$ interaction ChPT Lagrangian in \cref{eq:Leffphi} then leads to
\begin{align}
    \label{eq:LeffphiNLO}
    \lag_{\rm int}^\phi(U)
    \to 
    \lag_{\rm int}^\phi(U) + E_{14}\frac{\phi}{f}
    \langle \lambda_6 D_\mu U^\dagger D^\mu U\rangle
    \langle \chi_\phi' U^\dagger+{\rm h.c.}\rangle\,, 
\end{align}
where 
\beq
    \chi_\phi'
    =
    2B_0 {\rm diag}\left[(\kappa_{u}^\phi -2 \kappa_W+3K_\Theta)m_u,
    (\kappa_d^\phi-2 \kappa_W+3K_\Theta)m_d,(\kappa_s^\phi-2 \kappa_W+3K_\Theta)m_s\right] \, .
\eeq 
The dimensionless low energy constant $E_{14}$ in the $\cO(p^4)$ weak ChPT Lagrangian is estimated to be $\sim \cO(10^{-11})$ from fits of the $\cO(p^4)$ ChPT expressions to the $K\to \pi\pi$ and $K\to \pi\pi\pi$ data~\cite{Kambor:1991ah}.

%%%%%%%%%%%%%%%%%%%%%%%%%%%%%%%%%%%%
\section{Nonleptonic decays of light mesons involving $\phi$}
\label{sec:chiPT}
%%%%%%%%%%%%%%%%%%%%%%%%%%%%%%%%%%%%

Next, we calculate the $K \to \pi \phi$, $\eta^{(\prime)}\to \pi \phi$, and $\eta^\prime \to \eta \phi$ decay amplitudes based on the $\cO(p^2)$ chiral Lagrangian in \cref{eq:Leffphi}, including the partial $\cO(p^4)$ correction from interaction in \cref{eq:LeffphiNLO}.
The partial decay width for $M_1\to M_2\phi$ decay is given by
\begin{align}
    \Gamma(M_1\to M_2\phi)
=   \frac{1}{8\pi}\frac{p}{(m_1)^{2}} \big|\cM\big|^2 ,
\end{align}
with $m_{1,2}$ the masses of the two pseudoscalars, $M_{1,2}$, while $p=\big[\big(m_1^2-(m_2+m_\phi)^2 \big)\big(m_1^2-(m_2 - m_\phi)^2 \big)\big]^{1/2}/(2m_1)$ is the three-momentum of the outgoing particles in the $M_1$ rest-frame.
For the decay amplitudes, we work to partial NLO order in chiral expansion, $\cM=\cM_{\rm LO}+\cM_{\rm NLO}+\cdots$, with the results for $K \to \pi \phi$, $\eta^{(\prime)}\to \pi \phi$, and $\eta^\prime \to \eta \phi$ decays listed below. 

%%%%%%%%%%%%%%%%%%%%%%%%%%%%%%%%%%%%
\subsection{Kaon decays involving $\phi$}
\label{sec:KaonDecays}
%%%%%%%%%%%%%%%%%%%%%%%%%%%%%%%%%%%%

The $\Delta S=1$ interactions in \cref{eq:Leffphi} lead to $K^+\to \pi^+\phi$ and $K_{S,L}\to \pi^0\phi$ decays. 
At LO in chiral counting the $K^\pm\to \pi^\pm \phi$ amplitude is given by (the kaon and pion masses in the expressions below are understood to be for the charged states)
\beq
    \begin{split}
    \cM(K^+\to \pi^+  &\phi)_{\rm LO}
    = 
    \frac{1}{f} \bigg\{
     \frac{1}{2}(\kappa_W-K_\Theta)\left[\gamma_1
        \big(m_{K}^2-m_\phi^2+m_{\pi}^2\big)-2\gamma_2\big(m_K^2+m_\pi^2\delta_I\big)\right]   \\
    &+ \frac{1}{4}\kappa_{sd}^{\phi*}\left[2m_K^2-m_\pi^2(1-\delta_I)\right]+\frac{m_\pi^2}{4}\kappa_{ds}^{\phi*}(1+\delta_I)\\
    &-\frac{m_\pi^2}{4}\bigg(\kappa_u^\phi-\frac{\kappa_{s+d}^\phi}{2}\bigg)\gamma_1(1-\delta_I)\\
    &+\frac{m_\pi^2\kappa_{s-d}^\phi}{8(m_K^2-m_\pi^2)}\Big(m_K^2\big[(3+\delta_I)\gamma_1-4(1+\delta_I)\gamma_2\big]-m_\pi^2\big[(1-\delta_I)\gamma_1-2\gamma_2\big]\Big)
    \bigg\}\,,
\end{split}
\eeq
where $\delta_I\equiv (m_d-m_u)/(m_d+m_u)\simeq 0.37$~\cite{ParticleDataGroup:2024cfk} measures the isospin breaking effect due to $m_d\neq m_u$. 
We work to first order in $\delta_I$, and neglect corrections of order
$\cO(\delta_I^2)$. 
We have also shortened $\kappa_{s\pm d}^\phi\equiv\kappa_s^\phi \pm\kappa_d^\phi$. 
The NLO amplitude is given by 
\beq
    \label{eq:amplitudeKp:NLO}
    \cM(K^+\to \pi^+ \phi)_{\rm NLO}
    = 
    \frac{4}{f} \frac{E_{14}}{f_\pi^2}\Big(\kappa_{u}^\phi m_{\pi}^2 
    + \kappa_{s+d}^\phi m_{K}^2  
    \Big)\big(m_{K}^2-m_\phi^2+m_{\pi}^2\big) +\cdots,
\eeq
with the ellipses denoting the remaining NLO terms that we do not consider.  
In this expression, we also neglected subleading $\kappa_{s-d}^\phi,\kappa_W,K_\Theta$ contributions, and assumed the isospin limit, $\delta_I\to 0$.

In the expressions for the decays of neutral kaons, $K_{L,S}\to \pi^0 \phi$, we can ignore small CP-violating effects from the SM weak interactions and set $K_{L,S}=(K^0\pm\overline{K}^0)/\sqrt{2}$. 
The effect due to  $\pi^0-\eta/\eta'$ mixing is captured by a percent-level modification of the decay amplitude\footnote{For brevity $\pi_{\rm int}^0$ was denoted simply as $\pi^0$  in \cref{eq:Pi:explicit}.}
\beq
    \label{eq:pi-eta-eta':mixing}
    \cM(K_{L,S}\to \pi^0 \phi) 
    = 
    \left(1-R_{\pi^0-\eta/\eta'}\right)\mathcal{M}(K_{L,S}\to \pi_{\rm int}^0 \phi)\,,
\eeq
where $R_{\pi^0-\eta/\eta'}\approx 0.011$ (for analytical expression for $R_{\pi^0-\eta/\eta'}$, and the $\pi^0,\eta,\eta'$ mass matrix diagonalization, see \cref{sec:pietamixing}). 
The decay amplitudes at LO in chiral expansion are given by 
(the kaon and pion masses are now for those of neutral states)
\begin{align}
  \cM(K_L\to \pi_\text{int}^0\phi)_\text{LO}
    = 
    -  \frac{1}{f}\Bigg\{
    &   \frac{1}{2}(\kappa_W-K_\Theta)\left[\gamma_1
    \big(m_{K}^2-m_\phi^2+m_{\pi}^2\big)-2\gamma_2m_K^2\right]\\
    &+ \frac{1}{4}\text{Re}\, \kappa_{sd}^\phi\left[2m_K^2-m_\pi^2(1+\delta_I)\right]+\frac{m_\pi^2}{4}\text{Re}\,\kappa_{ds}^\phi(1+\delta_I)\\
    & +\frac{\kappa_{s-d}^\phi m_\pi^2}{4(m_K^2-m_\pi^2)}(\gamma_1-\gamma_2)\left[2m_K^2(1+\tilde\delta_I)-m_\pi^2\right]
    \Bigg\}\,,
\end{align}
with $\tilde \delta_I=\delta_I\big\{1+m_\pi^4/ [2 m_K^2(m_K^2-m_\pi^2)]\big\}\simeq 1.003 \delta_I$, and
\begin{align}
    \label{eq:KSpi0:LO}
    \cM(K_S\to \pi_{\rm int}^0 \phi)_\text{LO} 
    =
    \frac{1}{4f}\Bigg\{{\rm Im}\,\kappa_{sd}^\phi\left[2m_K^2-m_\pi^2(1+\delta_I)\right]+{\rm Im}\,\kappa_{ds}^\phi m_\pi^2(1+\delta_I)\Bigg\}\,,
\end{align}
At NLO we have 
\beq
    \begin{split}
    \label{eq:amplitudeKL:NLO}
    \cM(K_L\to \pi_\text{int}^0 \phi)_{\rm NLO}
    = 
    -\frac{4}{f} \frac{E_{14}}{f_\pi^2}\Big(\kappa_{u}^\phi m_{\pi}^2 + \kappa_{s+d}^\phi m_{K}^2  
    \Big)\big(m_{K}^2-m_\phi^2+m_{\pi}^2\big) +\cdots,
    \end{split}
\eeq
where as before we only display the dependence on the $\kappa_u^\phi$ and $\kappa_{s+d}^\phi$ parameters from the $E_{14}$ operator. 
This operator does not contribute to $\cM(K_S\to \pi^0 \phi)$ so that in the approximation we work with, we do not include any NLO correction to \cref{eq:KSpi0:LO}.
Note that the $K_S\to \pi^0\phi$ decay is CP-violating and thus the amplitude only depends on the imaginary part of the penguin-like flavor off-diagonal term, $\kappa_{sd,ds}^\phi$. 
Conversely, the $K_L\to \pi^0\phi$ decay preserves the CP symmetry and only the real part of $\kappa_{sd,ds}^\phi$ contributes to the amplitude.

In the above expression the $m_K$ and $m_\pi$ denote charged (neutral) kaon and pion masses, respectively, for $K^+$ ($K_{L,S}$) decays. 
In the isospin limit, we have $\pi^0_{\rm int}=\pi^0$, $m_{\pi^0}=m_{\pi^+}$, $m_{K^0}=m_{K^+}$ and $\kappa_u^\phi=\kappa_d^\phi$, implying the so-called Grossman-Nir relation among the kaon decay rates~\cite{Grossman:1997sk} (see also~\cite{Leutwyler:1989xj}),
\beq
    \cM(K^+\to \pi^+\phi)
    +
    \cM(K_L\to \pi^0\phi)+i\cM(K_S\to \pi^0\phi)=0\,.
\eeq
Furthermore, using~\cref{eq:cG:phi:Higgsmixed,eq:cgamma:phi,eq:kappapsi:phi:Higgsmixed} yields the same decay amplitude for $K^+\to\pi^+\phi$ and $K_L\to \pi^0\phi$ in the Higgs-mixed case, 
\beq
    \begin{split}
    \cM(K&\to \pi \phi)_{\text{Higgs--mix.}}=
    \\
    & 
    \frac{\sin\theta}{v}m_{K}^2 \left\{
     \frac{7}{18}\left[\gamma_1\left(1-\frac{m_\phi^2-m_{\pi}^2}{m_{K}^2}\right) - 2\gamma_2 \right]   
    + \frac{3G_F}{16\sqrt{2}\pi^2}
    \sum_{i=c,t}m_i^2 V_{is}V_{id}^\ast
    \right\}\,,
    \end{split}
\eeq
which agrees with~\cite{Leutwyler:1989xj}.

%%%%%%%%%%%%%%%%%%%%%%%%%%%%%%%%%%%%
\subsection{\texorpdfstring{Decays of $\eta,\,\eta'$}{eta, etaprime} mesons involving $\phi$}
\label{sec:EtaDecays}
%%%%%%%%%%%%%%%%%%%%%%%%%%%%%%%%%%%%

The $\eta/\eta'\to \pi^0 \phi$ decays are a result of QCD transition amplitudes supplemented by mixing of $\phi$ with the light mesons, giving 
\begin{align}
    \cM(\eta\to \pi^0_{\rm phys} \phi) 
    =& 
    \frac{m_\pi^2}{2f}\cos(\theta_{\eta\eta'}+\alpha)
    \left[K_I +K_8\frac{\delta_I(m_{\eta'}^2-m_\eta^2)m_\pi^2\sin^2(\theta_{\eta\eta'}+\alpha)}
    {(m_\eta^2-m_\pi^2)(m_{\eta'}^2-m_\pi^2)}\right] \,,\\
    \cM(\eta'\to \pi^0_{\rm phys} \phi) 
    =& 
    \frac{m_\pi^2}{2f}\sin(\theta_{\eta\eta'}+\alpha)
    \left[K_I-K_8\frac{\delta_I(m_{\eta'}^2-m_\eta^2)m_\pi^2\cos^2(\theta_{\eta\eta'}+\alpha)}
    {(m_\eta^2-m_\pi^2)(m_{\eta'}^2-m_\pi^2)}\right]\,, 
\end{align}
where the $\pi^0$--$\eta/\eta'$ mixing was treated at leading order in $\delta_I$ (see \cref{sec:pietamixing}), and we have defined 
\begin{align}
    \label{eq:KIdef} 
    K_I
    &\equiv 
    \kappa_d^\phi-\kappa_{u}^\phi+\delta_I(\kappa_{u}^\phi+\kappa_d^\phi-2 K_\Theta)\,,\\
    \label{eq:K8def}
    K_8
    &\equiv 
    \kappa_{u}^\phi+\kappa_d^\phi-2 r_s \kappa_s^\phi+2(r_s-1) K_\Theta\,,
\end{align} 
where $r_s$ is the ratio of quark masses, 
\begin{align}
    \label{eq:rs}
    r_s
    \equiv 
    \frac{2 m_s}{m_u+m_d} 
    = \frac{\sqrt{3}m_\eta^2\cos\theta_{\eta\eta'}-m_\pi^2\cos(\theta_{\eta\eta'}+\alpha)}{\sqrt{2}m_\pi^2\sin(\theta_{\eta\eta'}+\alpha)}= 26.8\,,
\end{align}
while the $\eta-\eta'$ mixing angle is $\theta_{\eta\eta'}\simeq 
 - 22^\circ$,  $\alpha=\arctan\sqrt 2$, and we derive  
$r_s = 26.8$ from the light quark masses \cite{ParticleDataGroup:2024cfk}. 
The  $\eta'\to \eta \phi$  decay amplitude is, similarly, given by 
\begin{align}
    \cM(\eta'\to \eta \phi)
    = 
    -\frac{m_\pi^2}{4f}\sin[2(\theta_{\eta\eta'}+\alpha)]
    \left[K_8+K_I\frac{\delta_I(m_\eta^2m_{\eta'}^2-m_\pi^4)}
    {(m_\eta^2-m_\pi^2)(m_{\eta'}^2-m_\pi^2)}\right].
\end{align}
In the above $\eta^{(\prime)}$ decay amplitudes we used the LO chiral Lagrangian relations between interactions and mass eigenstates (see \cref{sec:pietamixing}), setting $\theta_{\eta\eta'}$ and $m_{\pi}$, $m_{\eta}$, $m_{\eta'}$ to their measured values. This is not a unique choice --- other uses of measurements to fix numerical values in the LO chiral Lagrangian would lead to expressions for $\eta^{(\prime)}\to\pi\phi$ and $\eta'\to\eta\phi$ decay amplitudes that differ by terms that are of higher order in chiral counting (the origin of this numerical ambiguity can be traced to the fact that the LO chiral Lagrangian predictions cannot accommodate fully the measurements in the $\eta-\eta'$ system). 
The description can be systematically corrected by using higher order of the $p^2$ expansion (see {\it e.g.}~\cite{Gerard:2004gx}), which beyond our scope.

Note that the dependence on $K_\Theta$ always enters through $\kappa_q^\phi-K_\Theta$ linear combinations.
Furthermore, in the limit of infinite $\eta'$ mass, which implies  $\theta_{\eta\eta'}\to 0$ and $r_s\to (3m_\eta^2/m_\pi^2-1)/2$, the $\cM(\eta\to \pi^0_{\rm phys}\phi)$ amplitude vanishes for a  Higgs-mixed scalar~\cite{Ellis:1975ap}, up to NLO corrections in the chiral Lagrangian~\cite{Leutwyler:1989xj}.

The above decays can be used to place limits on the couplings of $\phi$ to the SM fermions. We first introduce in \cref{sec:UV} several UV models that can lead to a light scalar with general flavor structure, and then derive constraints in \cref{sec:bounds}.

%%%%%%%%%%%%%%%%%%%%%%%%%%%%%%%%%%%%
\section{UV models for general light scalar}
\label{sec:UV}
%%%%%%%%%%%%%%%%%%%%%%%%%%%%%%%%%%%%

Our main interest are scalars that are not the usual pNGBs, so that they can have flavor diagonal scalar couplings to the SM fermions (pNGBs due to spontaneous breaking of a global symmetry have derivative couplings, where derivative couplings of the form $\partial_\mu \phi \bar \psi_i \gamma^\mu \psi_i$ vanish after integration by parts). 
This means that the reason why the scalar $\phi$ is light will not be so straightforward. 
For most part we will bypass this question, but will construct an example where the scalar is naturally light in \cref{sec:dark:dilaton}.

A completely UV agnostic treatment of a general light scalar interactions with the SM model is accomplished by integrating out the heavy degrees of freedom, giving SMEFT${}_\phi$ effective field theory above the electroweak scale, and WET${}_\phi$ below it. 
The two EFTs have already been discussed in \cref{sec:SMEFTphi} and \cref{sec:EFT}, respectively. 
Further structure can be uncovered in concrete UV models~\cite{Batell:2021xsi}. 
Below, we discuss in \cref{sec:2HDM:phi} the limit of a light scalar in the context of a two-Higgs doublet model extended by a light singlet (2HDM${}_\phi$), \cref{sec:dark:dilaton} contains an example of a light dark dilaton, and \cref{sec:VLQ} an extension of the SM by a set of vector-like quarks and a light scalar (VLQ${}_\phi$). 

%%%%%%%%%%%%%%%%%%%%%%%%%%%%%%%%%%%%
\subsection{2HDM extended by a light singlet}
\label{sec:2HDM:phi}
%%%%%%%%%%%%%%%%%%%%%%%%%%%%%%%%%%%%

A simple UV complete example of a light scalar with general flavor structure is a two Higgs doublet model~(2HDM) extended by a singlet scalar $\phi$ (2HMD${}_\phi$)~\cite{Batell:2021xsi}. 
In the decoupling limit for the 2HDM~\cite{Gunion:2002zf}, the scalar sector at the weak scale consists of only two states, the SM-like Higgs $h$, with mass $m_h\simeq 125$\,GeV, and the light scalar $\phi$, while all the other scalars are heavier. 
In this case, the theory matches onto SMEFT${}_\phi$, which was discussed in \cref{sec:BSMint}. 

A qualitatively new behavior is obtained in the limit of alignment without decoupling for the 2HDM scalar potential~\cite{Carena:2013ooa,Craig:2013hca,Haber:2013mia,Haber:2018ltt}. 
It is convenient to work in the so called Higgs basis, in which the two Higgs doublets $H_{1,2}$ are chosen such that only the neutral component of the first field obtains a nonzero vacuum expectation value, $\langle H_1^0\rangle =v/\sqrt{2}$, $\langle H_2^0\rangle=0$, where $v=246\,$GeV.
Assuming  for simplicity a CP-conserving scalar potential, the CP-even Higgs squared matrix is  given by (see, \eg,~\cite{Haber:2018ltt})
\begin{align}
    \label{eq:MH:matrix}
    \cM_{H}^2
    =
    \begin{pmatrix}
        Z_1 v^2 & Z_6 v^2 \\
        Z_6 v^2 & m_A^2+Z_5 v^2
    \end{pmatrix}\,,
\end{align}
where $m_A$ is the mass of the CP-odd Higgs scalar, while the other terms come from the following terms in the 2HDM scalar potential 
\begin{align}
    V
    \supset 
    \frac{1}{2} Z_1(H_1^\dagger H_1)^2 
    +\Big\{ \frac{1}{2}Z_5 ( H_1^\dagger H_2)^2+Z_6 (H_1^\dagger H_1)(H_1^\dagger H_2)
    +{\rm h.c.}\Big\}.\,
\end{align}
The decoupling limit is obtained when $m_A^2+Z_5 v^2\gg Z_1 v^2$, while alignment without decoupling corresponds to the limit $|Z_6|\ll 1$, which is the limit we are interested in. 
Assuming for simplicity furthermore that $m_A^2+Z_5 v^2 >Z_1 v^2$ the two CP-even mass eigenstates, $h$ and $H$, are aligned with the $H_1^0$ and $H_2^0$ Higgs basis states, respectively.\footnote{If $m_A^2+Z_5 v^2 <Z_1 v^2$, the $h$ and $H$ mass eigenstates are instead aligned with $H_{2}^0$ and $H_1^0$, respectively, with trivial changes to our results.} 
In this limit, the Yukawa couplings of $h$ to the SM fermions $\psi$ are to a good approximation given by the SM Yukawas, $Y_\psi^h\simeq Y_\psi^{\rm SM}+\cO(Z_6 v^2/m_A^2)$. 
The couplings of $H$, $Y_\psi^H$, on the other hand, are completely arbitrary, only subject to experimental constraints such as the bounds on flavor changing neutral currents and bounds from direct searches. 

The couplings of the light singlet $\phi$ are induced from mass mixing with the two CP-even Higgses. 
Extending the mass matrix in \cref{eq:MH:matrix} to include also $\phi$ and diagonalizing it, we have
\begin{align}
    h=\sin\theta_h \phi +\cdots, 
    \qquad 
    H=\sin\theta_H \phi +\cdots,
\end{align}
where $h$ and $H$ are the mass eigenstates of $\cM_{H}^2$ as before, and ellipses denote other terms that do not include $\phi$. 
The Yukawa couplings of $\phi$ to the SM fermion $\psi$ are then given by
\beq
\begin{split}
\label{eq:2HDM:couplings}
    Y_\psi^\phi
    &=
    \sin\theta_h Y_\psi^h+\sin\theta_H Y_\psi^H 
    \\
    &\simeq 
    \sin\theta_h Y_\psi^{\rm SM}+\sin\theta_H Y_\psi^H \qquad \text{(alignment w/o decoupling)} \, .
\end{split}
\eeq
The first term in the second line in \cref{eq:2HDM:couplings} is the same as for the light Higgs-mixed scalar case. The deviations of $\phi$ couplings to the SM fermions from the light Higgs-mixed scalar limit are thus mainly due to $\phi$ mixing with the heavy CP-even scalar $H$.

The limits on flavor violating couplings of the heavy Higgs, $(Y_\psi^H)_{ij}, i\ne j$, are rather stringent and come from $K$--$\overline{K}$, $D$--$\overline{D}$, $B_{d,s}$--$\overline{B}_{d,s}$ mixing constraints for couplings to quarks and from rare lepton decays such as $\mu \to 3 e$, $\tau \to 3\mu$, etc, for couplings to leptons. 
Very approximately they are $(Y_\psi^H)_{ij} < r_{ij} \sqrt{(Y_\psi^{\rm SM})_{ii} (Y_\psi^{\rm SM})_{jj}} m_H^2/m_h^2$, with the numerical coefficient $r_{ij}$ ranging from ${\mathcal O}(10^{-3})$ for $(Y_\psi^H)_{e\mu, \mu e}$, to ${\mathcal O}(0.1)$  for $(Y_\psi^H)_{\tau\ell, \ell \tau}$ and $(Y_\psi^H)_{uc, cu}$, $(Y_\psi^H)_{sd, ds}$, $(Y_\psi^H)_{sb, bs}$, $(Y_\psi^H)_{db, bd}$,  and much less stringent for couplings involving top quark~\cite{Harnik:2012pb} (the exact bounds depend also on charged Higgs and CP-odd Higgs couplings -- for a recent analysis of 2HDM scenario where a single off-diagonal coupling to quarks is taken to be nonzero, see \cite{Kamenik:2023hvi}). 

An illustration of the general flavor-aligned light scalar limit can be obtained by taking $\theta_h\to0$, while $\theta_H\ne 0$ and only one of the diagonal $Y_\psi^H$ Yukawa couplings is nonzero.
While this is a rather ad-hoc limiting case, one could also contemplate other possible more motivated flavor structures for $h$ and $H$ Yukawa couplings that are also still compatible with the measurements. 
For instance, the third generation masses could come from the couplings to the SM-like Higgs, and the 1st and 2nd generation masses from the VEV of the heavier Higgs~\cite{Altmannshofer:2015esa,Ghosh:2015gpa,Altmannshofer:2016zrn}. 
We emphasize that in the 2HDM${}_\phi$ model there is plenty of freedom in the form of the couplings of $\phi$ to the first two generations of the SM fermions, especially for the diagonal couplings. 

%%%%%%%%%%%%%%%%%%%%%%%%%%%%%%%%%%%%
\subsection{Light dilaton from the dark sector}
\label{sec:dark:dilaton}
%%%%%%%%%%%%%%%%%%%%%%%%%%%%%%%%%%%%

Another example of a general flavor-aligned light scalar $\phi$ is a dark sector dilaton \cite{Salam:1969bwb,Isham:1970gz,Isham:1971dv,Ellis:1970yd,Ellis:1971sa,Low:2001bw}, i.e., a pNGB of a spontaneously broken conformal symmetry in the dark sector, where the conformal symmetry is also explicitly broken by small couplings to the visible sector (for the case where dark sector dilaton is the dark matter, see \cite{Arvanitaki:2014faa,Alachkar:2024crj,Appelquist:2024koa,Cyncynates:2024bxw,Hubisz:2024hyz,Redi:2020ffc,Hong:2022gzo}).  
Below the symmetry breaking scale $f$ the conformal symmetry is realized non-linearly, such that under scale transformations, $x^\mu\to e^{-\omega} x^\mu$, the dilaton undergoes a shift $\phi(x)\to \phi(x)+\omega f$. 

The mass of the dilaton is protected by approximate scale invariance. 
This is broken by quantum effects (RG running of couplings) and explicitly, at the classical level, by the presence of couplings of non-zero mass dimension. 
Both of these effects can be small, depending on the details of the conformal sector, which in this context means that they introduce dilaton mass that is well below the scale  of conformal symmetry breaking,  $m_\phi\ll 4\pi f$~\cite{Coradeschi:2013gda} (for a related discussion focused on Higgs being a light dilaton, see~\cite{Garriga:2002vf,Grinstein:2011dq,Goldberger:2007zk,Bellazzini:2013fga,Csaki:2007ns}). 
From now on, we will assume that indeed $m_\phi\ll 4\pi f$, so that the dilaton is the only dark sector state that is kinematically accessible in kaon decays.

Furthermore, we assume that there are interactions between dark sector and the visible sector. 
These necessarily break conformal invariance due to SM fermion and electroweak gauge boson masses being non-zero, though we assume that this breaking is small enough not to disrupt the conformal structure of the dark sector. 
The interaction of the light dark sector dilaton are thus of two types: {\em i)} the interactions entirely in the dark sector, $\cL_{\rm conf}$, which follow the structure dictated by the spontaneously broken conformal symmetry, and {\em ii)} the conformal symmetry breaking interactions with the SM fields, $\cL_{\text{SM}+\phi}$,
\beq
    \cL_\phi
    =
    \cL_{\rm conf}+\cL_{\text{SM}+\phi} \, .
\eeq 

In order to write down $\cL_{\rm conf}$ it is useful to introduce a conformal compensator field~\cite{Salam:1969bwb,Isham:1970gz,Isham:1971dv,Ellis:1970yd,Ellis:1971sa,Chacko:2012sy} $\chi(x)\equiv f e^{\phi(x)/f}$, which under the scale change transforms linearly, $\chi(x)\to e^{\omega} \chi(x)$. 
In terms of $\chi$ the low energy effective Lagrangian of a dilaton is given by~\cite{Chacko:2012sy}
\begin{align}
    \cL_{\rm conf}
    \supset 
    \frac{1}{2}  \partial_\mu \chi \partial^\mu \chi 
    +\frac{c}{\chi^4}\big(\partial_\mu \chi \partial^\mu \chi)^2
    - \frac{\kappa_0}{4!} \chi^4+\cdots\,,
\end{align}
where $c$ and $\kappa_0$ are dimensionless couplings. 
Note that dilaton has both derivative and non-derivative terms, in contradistinction to a pNGB of a spontaneously broken internal global symmetry, for which one can find a field redefinition such that it only has derivative couplings.  

In the absence of conformal symmetry violating effects the couplings of the dilaton to the massive SM fields would be such as to compensate the breaking of conformal invariance by the mass terms. The couplings of $\phi$ to the SM fermions would thus be proportional to the fermion masses~\cite{Chacko:2012sy}, 
\begin{align}
    \cL_{\text{SM}+\phi}
    \supset 
    \frac{\chi}{f}m_\psi \bar \psi \psi +\cdots
    =\phi \frac{m_\psi}{f}\bar \psi \psi +\cdots\,, 
\end{align}
where in the second equality we have expanded the conformal compensator field $\chi$. 
In this limit the light dark sector dilaton would have resembled the light Higgs-mixed scalar, but with the sine of the mixing angle, $\sin\theta$, replaced by $v/f$, in the interactions with the SM fields. 

However, for our purposes a more interesting case is when the small interactions between the dark sector and the visible sector explicitly break conformal invariance.
In this case the interactions between the light dark sector dilaton and the SM fermions can have a completely generic flavor structure,
\begin{align}
    \cL_{\rm EFT}
    \supset  
    c_{ij}^\psi {\chi} \bar \psi_i \psi_j +{\rm h.c.} +\cdots
    =c_{ij}^\psi \phi  \bar \psi_i \psi_j  +{\rm h.c.} +\cdots\,, 
\end{align}
where $c_{ij}^\psi $ are dimensionless complex couplings, assumed to be small, $|c_{ij}^\psi|\ll 1$. 
While from the bottom-up perspective the flavor structure of $c_{ij}^\psi $ is arbitrary, this may not be the case if the flavor structure of the SM itself is due to some underlying dynamics \cite{Altmannshofer:2022aml}. 
In \cref{sec:limiting:cases}  we were most interested in the limiting case, where  the $c_{ij}^\psi$ are almost flavor diagonal, but also differ from the predictions of the light Higgs-mixed scalar, $c_{ij}^\psi\ne \sin\theta_h y_i^{\rm SM} \delta_{ij}$. 
This limit of a general flavor-aligned light scalar is certainly possible to achieve in the light dark sector dilaton model, however, other flavor structures are in general also allowed. 
The light dark sector dilaton would in general also couple to the SM gauge bosons, i.e., $c_G^\phi, c_\gamma^\phi\ne 0$ in \cref{eq:Lint2GeV} in \cref{sec:EFT}. 
In our case, where the dark-sector interactions with the SM break conformal invariance, these couplings are free parameters from the low energy perspective.    

%%%%%%%%%%%%%%%%%%%%%%%%%%%%%%%%%%%%
\subsection{Vector-like quarks and a light scalar}
\label{sec:VLQ}
%%%%%%%%%%%%%%%%%%%%%%%%%%%%%%%%%%%%
 
As the last example of a UV complete model, which contains the general flavor-aligned light scalar, we consider the case, where the SM is supplemented by a light SM gauge singlet scalar, $\phi$, and by a set of VLQs carrying the following SU(3)$_c\times$SU(2)$_L\times$U(1)$_Y$ charges
\begin{align}
    U_L^i,U_R^i 
    \sim (\mathbf{3},\mathbf{1},2/3)\,,
    \quad\quad 
    D_L^i,D_R^i
    \sim (\mathbf{3},\mathbf{1}, -1/3)\,.
\end{align}
Above, the subscript $L(R)$ denotes a left-(right-)handed Weyl fermion field, while $i,j=1,2,3$ are generation indices (for simplicity we limit the discussion to the case of three generations of VLQs).
The VLQ interactions with the $\phi$ and the SM fields are given by the following renormalizable Lagrangian,
\beq
    \begin{split}
    \label{eq:LVLQ}
    \cL_{\rm VLQ}
     =& 
    \bar U(i\slashed{D}-M_U)U 
    + \bar D(i\slashed{D}-M_D)D  -\bar Q_L \hat y_U\widetilde H U_R
    -\bar Q_L \hat y_D H D_R
    \\
    & - \bar U_L \hat \lambda_U u_R \phi
    - \bar D_L \hat \lambda_D d_R \phi + {\rm h.c.}\,, 
    \end{split}
\eeq
where $M_{U,D}$ are $3\times3$ mass matrices for VLQs, $\hat y_{U,D}$ are $3\times 3$ dimensionless Yukawa couplings between the left-handed SM doublets $Q_L^i$ and the VLQs, and $\hat y_{U,D}$ the dimensionless $3\times 3$ couplings between VLQs, the right-handed SM fields, $u_R^i, d_R^i$ and $\phi$. 
Note that gauge invariance also allows mass mixing operators between the VLQs and the right-handed SM fields of the form, $\bar U_L M_U' u_R$ and  $\bar D_L M_D' d_R$.
These mass terms can be removed without loss of generality by an appropriate field redefinition of $U_R, u_R$ and $D_R, d_R$ fields (that is, by definition the SM right-handed quarks are the fields that have no mass terms with the left-handed VLQ fields $U_L, D_L$, after appropriate unitary transformations). 
Without loss of generality we can also work in the basis in which VLQ masses are diagonal, $M_{U,D}^{ij}=M_{U,D}^i\delta^{ij}$,

To simplify the results, we furthermore assume flavor diagonal $\phi$ interactions, $\hat \lambda_{U,D}^{ij}=\lambda_{U,D}^i\delta^{ij}$, while for Yukawa-like interactions with the Higgs, we use a Minimal Flavor Violating~(MFV)--like ansatz,
\begin{align}
    \label{eq:yUhat}
    \hat y_U^{ij}
    =
    \delta^{ij}y_U^j\,,
    \quad \quad
    \hat y_D^{ij}
    = 
    V^{ij}y_D^{\,j}\,,
\end{align}
where $V$ is the CKM matrix. 
The above flavor ansatz differs from the strict MFV limit, in that the $y_{U,D}^j$ do not equal the SM Yukawa couplings. 
Furthermore, there is an additional non-MFV source of flavor breaking, the $\lambda_{U,D}^i$ couplings, multiplying the interaction terms between $\phi$ and the mixed VLQ---SM-quark scalar currents in \cref{eq:LVLQ}. 
For simplicity, we assume that the $\lambda_{U,D}^i$ are diagonal in the quark mass basis, i.e., that they are aligned with the quark mass eigenstates. 
Once VLQs are integrated out, this then reproduces the alignment limit of the effective operators in \cref{eq:aligned:scalar}.

After EW symmetry breaking, the terms proportional to $\hat y_{U,D}$ in \cref{eq:LVLQ} induce mixings between the VLQs and the SM quarks. 
Because of the assumed flavor alignment each SM quark flavor mixes with just a single VLQ, so that the mass eigenstates follow from the redefinitions 
\begin{align}
    \begin{pmatrix} f \\ F \end{pmatrix} 
    \to 
    \begin{pmatrix}
        \cos\theta_f & \sin\theta_f \\ 
        -\sin\theta_f & \cos\theta_f
    \end{pmatrix} 
    \begin{pmatrix} f \\ F \end{pmatrix} \, ,
\end{align}
where $f$ denotes either one of the LH and RH SM quarks and $F$ its VLQ partner.
Assuming $\langle \phi\rangle = 0$, the LH and RH mixing angles are, respectively, given by
\begin{align}
    \tan 2\theta_{f_L}
    =
    \frac{\sqrt{2}y_FvM}{M^2-(y_f^2 +y_F^2)v^2/2}\,,
    \quad \quad
    \tan 2\theta_{f_R}
    =
    \frac{y_Fy_fv^2}{M^2-(y_f^2-y_F^2)v^2/2}\,,
\end{align}
where $y_f$ is the corresponding SM Yukawa coupling, and we dropped generation indices to shorten the expressions. Note that the phases of the VLQ fields can be adjusted so that the mixing angles are always real and positive. 
Below, we will assume the heavy VLQ limit, $M\gg y_{F} v, y_f v$, and work to leading order in the mixing angles,  $\theta_{f_L,}\theta_{f_R}\ll 1$,  
\begin{align}
    \theta_{f_L}
    \simeq 
    \frac{y_Fv}{\sqrt{2}m_F}\simeq\frac{y_F}{y_f}\frac{m_f}{M}\,,
    \qquad \qquad
    \theta_{f_R} 
    \simeq 
    \frac{m_f}{m_F}\theta_{f_L}\simeq\frac{m_f}{M}\theta_{f_L}\,,
\end{align}
where $m_f\simeq y_fv/\sqrt{2}$ and $m_F\simeq M$ are the SM quark and VLQ partner masses, respectively. 

Integrating out the VLQs gives SMEFT${}_\phi$ and then, after electroweak symmetry breaking, WET${}_\phi$. 
The tree level WET${}_\phi$ Wilson coefficients for coupling of $\phi$ to the fermion currents are given by (no summation over repeated indices and we identify the scale $f$ in \cref{eq:Lint2GeV} with the VLQ mass, $f=M$)
\beq
    \kappa_{u_i}^\phi
    = -\lambda_U^i \frac{y_U^i}{y_{u_i}},
    \qquad\qquad 
    \kappa_{d_i}^\phi
    = - \lambda_D^i \frac{y_D^i}{y_{d_i}}.
\eeq
The couplings to $W^\pm$, \cref{eq:LSMEFT:phi:gauge}, similarly vanish at tree level, $\kappa_W=c_W=0$.  
At low energies the $\phi$ couplings to gluons and photons arise mainly from integrating out the heavier quarks, and are thus given by \cref{eq:cG:phi,eq:cgamma:phi}, setting $\kappa_\tau=0$ in these expressions.

In the numerical examples below, we consider four illustrative benchmark cases: 
a single VLQ of mass $M$ mixing with either the SM $u,c$ or $t$ quark (denoted as ``$u$--only'', ``$c$--only'' and ``$t$--only'' benchmarks, respectively); and the example of 3 degenerate VLQs of common mass $M$, mixing with the $u,d,s$ quarks, where the magnitude of the mixing is proportional to the corresponding quark mass (the ``$uds$ model'' benchmark). That is, at tree level the nonzero SMEFT${}_\phi$ coefficients at $\mu\simeq M$ for the four benchmarks are 
\begin{align}
    \label{eq:KWtree:SMEFT}
    \text{$u(c,t)$--only:} \qquad &  
    \kappa_{u(c,t)}^\phi\ne0\,, \\
    \text{$uds$ model:} \qquad & \kappa_{u}^\phi=\kappa_{d}^\phi=\kappa_{s}^\phi\equiv \kappa^\phi\ne 0,
\end{align}
where, as in \cref{eq:kappa:ctphi:WET}, we adopted the suggestive WET${}_\phi$-like notation for the SMEFT${}_\phi$ coefficients, $\kappa_{u(c,t)}^\phi=\kappa_{11(22,33)}^u$, $\kappa_{s}^\phi=\kappa_{22}^d$. 

At $\mu=2$GeV the nonzero WET${}_\phi$ Wilson coefficients also include couplings to photons and gluons, generated at one loop when the heavy SM quarks get integrated out at their respective thresholds. 
Written in terms of the linear combinations $K_{I,8}$, \cref{eq:KIdef,eq:K8def}, and using the $K_\Theta$ normalization for couplings to gluons, \cref{eq:KTheta:main:text}, the flavor diagonal WET${}_\phi$ Wilson coefficients for these benchmarks are given by, 
\begin{align}
    \label{eq:KWtree}
    \text{$u$--only:} \qquad &  K_I
    =-(1-\delta_I)K_8 
    = -(1-\delta_I)\kappa_u^\phi\,,\quad
    K_\Theta=c_\gamma^\phi =0\,, \\
    \text{$c(t)$--only:} \qquad  &  - \frac{K_I}{2\delta_I}
    =\frac{K_8}{2(r_s-1)}
    =K_\Theta=\frac{\pi}{3\alpha}c_\gamma^\phi
    =\frac{2}{27}\kappa_{c(t)}^\phi
    \left(1+\frac{m_{c,t}^2}{M^2}\right)\,, \\
    \label{eq:uds:model}
    \text{$uds$ model:} \qquad & \frac{K_I}{\delta_I}=\frac{K_8}{1-r_s}=2\kappa^\phi\,,\quad  K_\Theta=c_\gamma^\phi =0\,,
\end{align}
where in all cases we set $m_{u,d,s}=0$ and have used that $\kappa_{b}^\phi=\kappa_W=0$.

%%%%%%%%%%%%%%%%%%%%%%%%%%%%%%%%%%%%
\begin{figure}
\begin{center}
    \includegraphics[height=10cm]{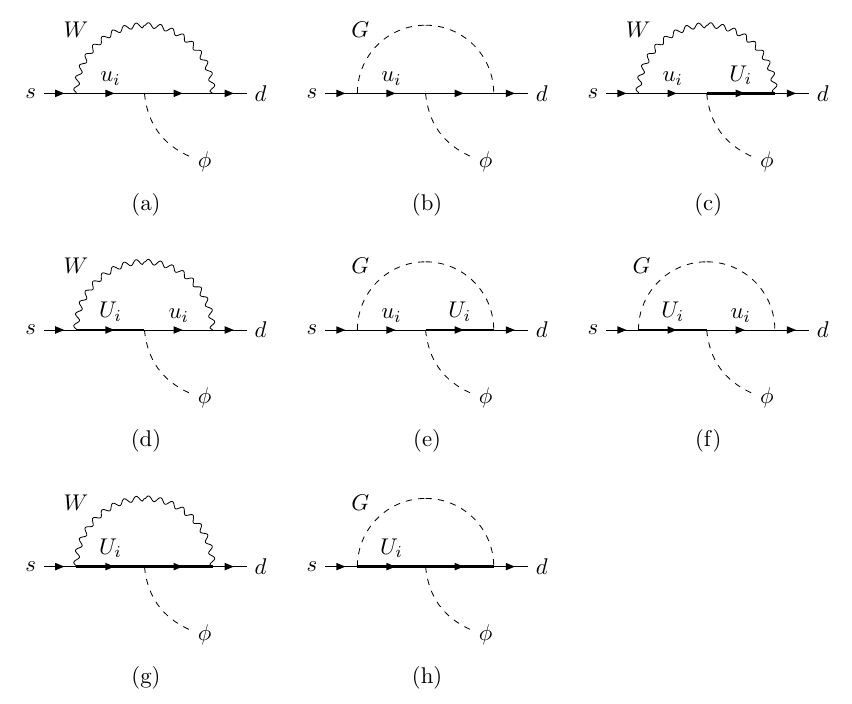}~
    \caption{One-loop penguin diagrams for $s\to d\phi$ transition in the VLQ model, assuming couplings of the SM fermions to only up-type VLQ $SU(2)_L$ singlets.}
    \label{fig:penguin-up}
\end{center}
\end{figure}
%%%%%%%%%%%%%%%%%%%%%%%%%%%%%%%%%%%%

\begin{figure}
\begin{center}
    \includegraphics[height=9cm]{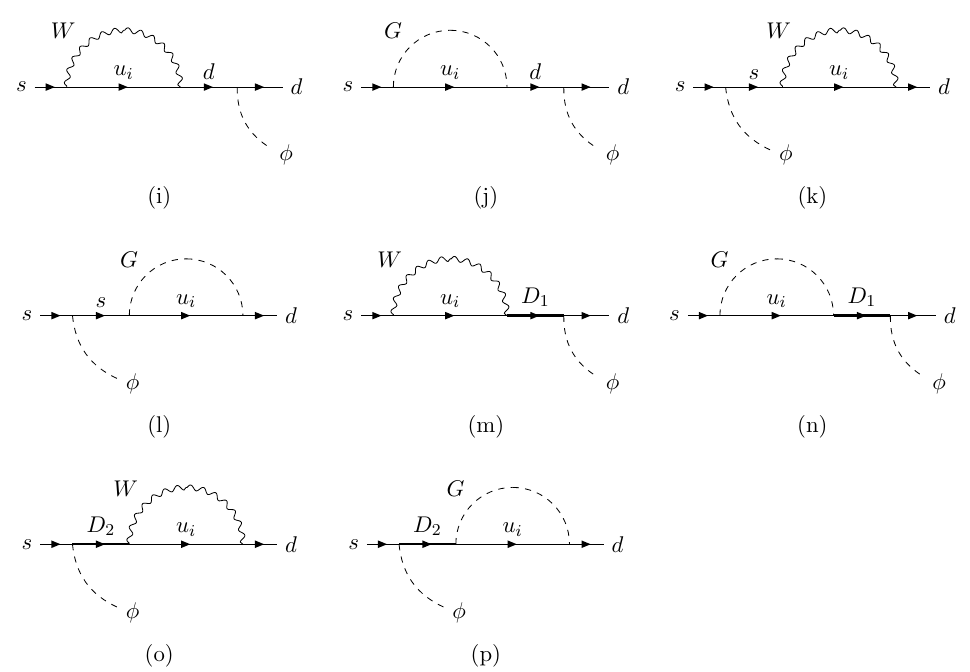}~
\caption{One-loop penguin diagrams for $s\to d\phi$ transition in the VLQ model, assuming couplings of the SM fermions to only down-type VLQ $SU(2)_L$ singlets.}
\label{fig:penguin-down}
\end{center}
\end{figure}

At one loop the $s\to d\phi$ couplings $\kappa_{ds}^\phi$ and $\kappa_{sd}^\phi$ get induced from penguin diagrams with heavy VLQs. 
These contributions are essential, in order to cancel the UV divergence from the SM quark loop diagrams in the SMEFT${}_\phi$ calculation. 
The resulting finite pieces are suppressed by the GIM mechanism.
We perform the calculation in the Feynman--{}'t\,Hooft gauge ($\xi=1$ in $R_\xi$ gauge) and  use dimensional regularization in $d=4-2 \epsilon$ dimensions  to regularize UV divergences.
The relevant Feynman rules are summarized in \cref{sec:Feynman-rules}. 
\Cref{fig:penguin-up} and \cref{fig:penguin-down} show the relevant penguin diagrams that involve $\phi$ couplings to the up- and down-type quarks, respectively. 
The contributions to the $\kappa_{sd}^\phi$ parameter in \cref{eq:Q12L} from each diagram with $\phi$ couplings to up-type quarks are 
\begin{align}
    \kappa_{sd}^{\phi, \,(a)}
    =& -\sum_i \kappa_{u_i}^\phi\zeta_{sd}^i \left[\frac{x_{i}^2-1-2x_{i}\log x_{i}}{(x_{i}-1)^3}\right]\,, \\
     \kappa_{sd}^{\phi, \,(b)}
    =& -\sum_i \kappa_{u_i}^\phi\zeta_{sd}^i\left[C_{\rm UV}-\frac{1}{2}-\frac{x_{i}^2-4x_{i} +3 +2x_{i}(x_{i}^2 -3x_{i}+3)\log x_{i}}{2(x_{i}-1)^3}\right]\,,\\
   \kappa_{sd}^{\phi, \,(c)+(d)}
    =& \sum_i \kappa_{u_i}^\phi\zeta^i_{sd}\left[\frac{x_{U_i}^2\log x_{U_i}}{(x_{U_i}-x_i)(x_{U_i}-1)^2}-\frac{x_i^2\log x_i}{(x_{U_i}-x_i)(x_i-1)^2}+\frac{1}{(x_{U_i}-1)(x_i-1)}\right]\,,\\
    \begin{split}
     \kappa_{sd}^{\phi, \,(e)+(f)}
    =&\sum_i \kappa_{u_i}^\phi\zeta^i_{sd}\left[C_{\rm UV} -\frac{1}{2} -\frac{x_i\log x_i}{2(1-x_i)^2(x_{U_i}-x_i)}[x_i(4-3x_i)-x_{U_i}(x_i-2)]\right. \\
    &\left.+\frac{x_{U_i}\log x_{U_i}}{2(1-x_{U_i})^2(x_{U_i}-x_i)}[x_{U_i}(4-3x_{U_i})-x_i(x_{U_i}-2)]\right.\\
    &\left.+\frac{3(1+x_i x_{U_i})-2(x_i+x_{U_i})}{2(x_{U_i}-1)(x_i-1)}\right]\,,
    \end{split}
\end{align}
and $\kappa_{sd}^{\phi \,(g),(h)}\simeq \cO(\theta_{u_L^i}^2)$ where $x_{i}\equiv m_{u_i}^2/m_W^2$,  $x_{U_i}\equiv m_{U_i}^2/m_W^2\simeq M^2/m_W^2$, $C_{\rm UV}\equiv 1/\epsilon-\gamma_E+\log(4\pi\mu^2/m_W^2)$ and 
\begin{align}
    \zeta_{sd}^i 
   \equiv \frac{G_F m_W^2}{4\sqrt 2 \pi^2}  x_i  V_{is}V_{id}^*.
\end{align}
Each diagram yields an equal contribution to the parameter $\kappa^\phi_{ds}$ in \cref{eq:Q12L}, which describe the interaction with opposite chirality.
%%%%%%%%%%%%%%%%%%%%%%%%%%%%%%%%
%%%%%%%%%%%%%%%%%%%%%%%%%%%%%%%%

The contributions to the parameter $\kappa_{ds}^\phi$ from each diagram with $\phi$ couplings to down and strange quarks are (neglecting $\cO(m_s^2/m_W^2)$ and $\cO(m_d^2/m_s^2)$ corrections),
\begin{align}
    \kappa_{ds}^{\phi, \,(i)}
    =& 
    \kappa_d^\phi\sum_i \zeta^{i}_{sd}\left[\frac{1}{x_i-1}-\frac{x_i\log x_i}{(x_i-1)^2}\right]\,,\\
    \kappa_{ds}^{\phi, \,(j)}
    =& 
    -\kappa_d^\phi\sum_i \zeta^{i}_{sd}\left[\frac{C_{\rm UV}}{2}+\frac{x_i-3}{4(x_i-1)}-\frac{x_i(x_i-2)\log x_i}{2(x_i-1)^2}\right]\,,\\
    \kappa_{ds}^{\phi,\,(k)}
    =& -
    \kappa_s^\phi\sum_i \zeta^i_{sd}\left[\frac{1}{x_i-1}-\frac{x_i\log x_i}{(x_i-1)^2}\right]\,,\\
    \kappa_{ds}^{\phi,\,(l)}
    =& 
    \kappa_s^\phi\sum_i \zeta^i_{sd}\left[\frac{3C_{\rm UV}}{2}+\frac{5x_i-7}{4(x_i-1)}-\frac{x_i(3x_i-4)\log x_i}{2(x_i-1)^2}\right]\,,\\
    \kappa_{ds}^{\phi,\,(n)}
    =& 
    -\kappa_d^\phi\sum_i \zeta^{i}_{sd}\left[C_{\rm UV}+1-\frac{x_i\log x_i}{x_i-1}\right]\,,
\end{align}
and $\kappa_{ds}^{\phi,\, (m)}\simeq\cO(m_s^2/M^2)$, $\kappa_{ds}^{\phi,\, (o),(p)}=0$. 
When $\kappa^\phi_d = \kappa_s^\phi$, the total contribution is $\kappa_{ds}^\phi \simeq 0$.

The contributions to the opposite chirality parameter, $\kappa_{sd}^\phi$, are
\begin{align}
    \kappa_{sd}^{\phi,\, (l)}=-
    \kappa_{sd}^{\phi,\, (p)}
    =& \kappa_s^\phi\sum_i \zeta_{sd}^i\left[C_{\rm UV}+1-\frac{x_i\log x_i}{x_i-1}\right]\,,
\end{align}
and $\kappa_{sd}^{\phi,\,(i),(j),(k)}\simeq\cO(m_d^2/m_s^2)$, $\kappa_{sd}^{\phi,\, (o)}\simeq \cO(m_d^2/M^2)$, $\kappa_{sd}^{\phi,\,(m),(n)}=0$, such that the total contribution is $\kappa_{sd}^\phi\simeq 0$.   

In the up-sector, the UV divergence  of the SM quark loop diagram in~\cref{fig:penguin-up}(b) is canceled by the divergence of the VLQ loop diagrams in ~\cref{fig:penguin-up}(e) and (f). Conversely, in the down-sector, only the UV divergence of the SM diagram in \cref{fig:penguin-down}(l) is canceled by the divergence of the VLQ diagram in~\cref{fig:penguin-down}(p), making $\kappa_{sd}^\phi$ finite. The remaining divergences in $\kappa_{ds}^\phi$ only vanish for $\kappa_d^\phi=\kappa_s^\phi$. Away from this limit, non-zero counter-terms must be introduced, as explained in \cref{sec:renormalization}.   

Summing the diagrams in~\cref{fig:penguin-up}, the full penguin contribution for $\phi$ that couples only to $i=u,c,t$ quarks (and their VLQ partners) is given by
\begin{align}
    f_i^{{\rm VLQ}}(x_i,x_{U_i})
    =
    1-\log x_{U_i}-\frac{2}{(x_i-1)^2}
    +\frac{x_i (4 - 3 x_i + x_i^2) \log x_i}{(x_i-1)^3}\,,
\end{align}
where we assumed $x_{U_i}\gg 1$ for the VLQ contribution. 
For the up and charm quarks the power suppressed terms can be ignored, so that the loop function simplifies to $f_i^{{\rm VLQ}}(0,x_{U_i})=-1-\log x_{U_i}$.
Summing the contributions, we have finally for the four benchmark cases, 
\begin{align}
\text{$u$--only, $uds$ model:}\qquad \kappa_{sd}^{\phi}
    =\kappa_{ds}^\phi=&0,
    \\
\text{$c$--only:}\qquad    \kappa_{sd}^{\phi}
    =\kappa_{ds}^\phi=& - \frac{3 G_F}{8\sqrt 2\pi^2}\kappa_c^\phi m_c^2 
    V_{cs}V^*_{cd}\left[\log\left(\frac{M^2}{m_W^2}\right)+1\right],
    \\
    \begin{split}
\text{$t$--only:}\qquad   \kappa_{sd}^{\phi}
    =\kappa_{ds}^\phi=& - \frac{3 G_F}{8\sqrt 2\pi^2}\kappa_t^\phi m_t^2 V_{ts}V^*_{td} \biggr[
     \log\left(\frac{M^2}{m_W^2}\right)-1
     \\
     &\quad+\frac{2}{(x_t-1)^2}-\frac{x_t(4-3x_t+x_t^2)\log x_t}{(x_t-1)^3}\biggr].
     \end{split}
\end{align}
%

%%%%%%%%%%%%%%%%%%%%%%%%%%%%%%%%%%%%
\section{Phenomenology of a general flavor aligned scalar}
\label{sec:bounds}
%%%%%%%%%%%%%%%%%%%%%%%%%%%%%%%%%%%%

Next, let us turn to the phenomenology of a general flavor-aligned light scalar, limiting the discussion to $\phi$ that is light enough so that it can be produced in kaon, $\eta$ and $\eta'$ decays.
The main goal of this section is to use the experimental data on such rare decays to set bounds on low-energy effective parameters, $\kappa_{u,d,s}^\phi$, $K_\Theta$ and $\kappa_W$, as well as on the parameters in a sample UV model, for which we choose the singlet VLQ model introduced in \cref{sec:VLQ}.
We limit the discussion to representative cases of constraints, see, \eg, Ref.~\cite{Gan:2020aco} for a more detailed analysis of rare $\eta$ and $\eta'$ decays.

The preferred parametrization of low-energy couplings is  $\kappa_W-K_\Theta$, $\kappa_{s- d}^\phi$, $\kappa^\phi_u-\kappa_{s+d}^\phi/2$, $\kappa_{sd,ds}^\phi$, and $K_I$, $K_8$, which is the one we use to report the constraints. 
That is, to the order we work the $\eta/\eta'\to \pi^0\phi$ decay amplitudes, \cref{sec:EtaDecays}, are functions of just the $K_I$ and $K_8$ combinations of tree-level parameters, see \cref{eq:KIdef,eq:K8def}. 
The $K\to \pi \phi$ decay amplitudes depend on $\kappa_W-K_\Theta$, $\kappa_{s-d}^\phi$, $\kappa^\phi_u-\kappa^\phi_{d+s}/2$ and $\kappa_{sd}^\phi$ at LO in chiral counting, while at (partial) NLO the $E_{14}$ amplitude also introduces dependence on $\kappa_u^\phi$ and $\kappa_{s+d}^\phi$. 
Since we do not perform a complete NLO analysis, we use these partial NLO expressions merely to obtain a rough estimate of bounds for the $uds$-model VLQ benchmark, \cref{eq:uds:model}.

The four VLQ model benchmarks highlight the drastically different limits on new physics that are obtained in various limiting cases of the assumed flavor structure. 
The $uds$ model benchmark, for instance, demonstrates that a large suppression of $K\to \pi \phi$ bounds is possible, since in this limit the dominant contribution arises only from $\cO(p^4)$ operators in the weak chiral Lagrangian. 
Out of the four benchmarks the $uds$ model benchmark is thus the one where kaon decays are the least constraining.  
For the case where the $\phi$ scalar has sizable couplings to the top quark, the $K\to \pi \phi$ decay amplitudes are dominated by the penguin contribution $\kappa_{sd}^\phi$. 
As a result of this, the $K\to \pi \phi$ searches place the most stringent constraints on the  $t$--only benchmark, out of all four benchmarks.  
The charm-quark-only and up-quark-only benchmarks sit somewhere between the above two limiting cases. 
A moderately suppressed penguin contribution $\kappa_{sd}^\phi \propto m_c^2/m_W^2$ for the charm-quark-only benchmark is comparable to the $K_\Theta$ contribution. 
This then results in significantly more stringent $K\to \pi \phi$ constraints, then one obtains for the up-quark only benchmark, for which  the $\kappa_{sd}^\phi$ coupling is highly suppressed. 

The phenomenology of a general flavor aligned light scalar also depends strongly on the decay modes of $\phi$. 
To simplify the analysis,  we consider two limiting scenarios: 
{\em i}) the case where $\phi$ decays predominantly to photons, {\it i.e.}  BR$(\phi\to\gamma\gamma)\approx1$, and {\em ii)} the case where $\phi$ decays to additional dark sector states, which may be absolutely stable, or simply appear as invisible on the scales of the experiment, {\it i.e.} BR$(\phi\to\text{invisible})\approx1$.

%%%%%%%%%%%%%%%%%%%%%%%%%%%%%%%%%%%%
\subsection{Light scalar decaying to photons}
\label{sec:phigammagamma}
%%%%%%%%%%%%%%%%%%%%%%%%%%%%%%%%%%%%

%%%%%%%%%%%%%%%%%%%%%% 
\begin{table}[t]
\centering
\begin{tabular}{ccc}
    \hline\hline
    decay mode & data & SM prediction   \\
    \hline
    $\eta\to \pi^0\gamma\gamma$ 
    & $(2.55\pm0.22)\times 10^{-4}$~\cite{ParticleDataGroup:2024cfk} 
    & $(1.8\pm 0.5)\times 10^{-4}$ \cite{Gan:2020aco}   \\
    $\eta^{\prime}\to \pi^0\gamma\gamma$ 
    & $(3.20\pm0.24)\times 10^{-3}$~\cite{ParticleDataGroup:2024cfk}  
    & $(2.91\pm 0.21)\times 10^{-3}$ 
    \cite{Escribano:2018cwg}  \\
    $\eta^\prime\to\eta\gamma\gamma$ 
    & $(8.25 \pm 3.48) \times 10^{-5}$~\cite{BESIII:2019ofm}
    &  $(1.17\pm 0.08)\times 10^{-4}$ 
    \cite{Escribano:2018cwg} \\
    \hline
    $K^+\to \pi^+ \phi, \phi \to\gamma\gamma$ 
    & bump-hunt for $m_\phi\gtrsim 0.2\,\GeV$~\cite{NA62:2023olg}  
    & data driven~\cite{NA62:2023olg,Ecker:1987hd,DAmbrosio:1996cak}   \\
    $K_{L}\to\pi^0\gamma\gamma$ 
    & $(1.27\pm0.03)\times 10^{-6}$~\cite{ParticleDataGroup:2024cfk}
    & data driven~\cite{NA48:2002xke,KTeV:2008nqz}   \\
    $K_{S}\to\pi^0\gamma\gamma$ 
    & $(4.9\pm1.8)\times 10^{-8},\, z>0.2$~\cite{NA48:2003ydp}
    &  $3.8\times 10^{-8},\, z>0.2$~\cite{Gerard:2005yk,Cirigliano:2011ny}   \\
    \hline\hline
\end{tabular}
\caption{A list of relevant data (2nd column) on branching ratios, $\text{BR}(M_1\to M_2\gamma\gamma)$,   with the SM predictions listed in the 3rd column. For $K^+\to\pi^+\gamma\gamma$ and $K_{S}\to\pi^0\gamma\gamma$ channels, $z \equiv m_{\gamma \gamma}^2/m_{K}^2 > 0.2 $ is imposed in order to eliminate the overwhelming  SM background from $K^+ \to \pi^+ \pi^0$ and $K_S \to \pi^0 \pi^0$ decays, respectively. 
The SM predictions for 
$K^+\to\pi^+\gamma\gamma$ and $K_{L}\to\pi^0\gamma\gamma$ are fits to $\mathcal{O}(p^6)$ ChPT expressions (see also main text).}
\label{tab:dataSMpred}
\end{table}
%%%%%%%%%%%%%%%%%%%%%% 

Let us first consider the case where $\phi$ is assumed to decay promptly to photons, and that this is the dominant decay mode. 
We then use the $M_1\to M_2 \gamma\gamma$ decay rates listed in \cref{tab:dataSMpred}, along with their experimental values and the SM predictions, to constrain the low energy interactions of the light scalar $\phi$. 
Comparing the measured radiative $\eta^{(\prime)}$ decay rates with the SM predictions one can derive bounds on $M_1\to M_2 (\phi\to \gamma\gamma)$ decays for $\phi$ mass lighter than  $m_\phi < m_{\eta^{(\prime)}}-m_\pi$.  
The $\phi$ interactions can also be constrained by the $K\to\pi \gamma\gamma$  transitions: from the bump-hunt in the $\gamma\gamma$ invariant mass spectrum in $K^+\to\pi^+ \gamma\gamma$ decays~\cite{NA62:2023olg}, and from the 
comparison of the SM prediction for BR$(K_S\to \pi^0\gamma\gamma)$ at ${\mathcal O}(p^4)$  in ChPT expansion with the experimental results. In both cases the $\phi$ masses are required to be above  $m_\phi^2>0.20\,m_K^2$ because of the $K^+
\to\pi^+\pi^0$ and $K^0\to\pi^0\pi^0$ backgrounds. No useful bounds can be obtained from the integrated branching ratios BR$(K^+\to\pi^+ \gamma\gamma)$ and BR$(K_L \to\pi^0 \gamma\gamma)$, since the ${\mathcal O}(p^4)$ predictions lie significantly below the experimental measurements, while the ${\mathcal O}(p^6)$ corrections are then fit to the two measured branching ratios, providing no additional predictive power. Note also, that no bump hunt search has yet been performed on $\gamma\gamma$ spectrum in $K_L\to \pi^0\gamma\gamma$ decay, though, this would be a useful constraint in the future. 

For $m_\phi\ll m_{\eta^{(\prime)}}-m_\pi$, the $\eta$ and $\eta'$ decays set the following 95\,\%~confidence level~(CL) interval upper bounds on the $K_I$ and $K_8$ parameters,  
\begin{align}
    \abs{K_I} < 2.7 
    \, , \qquad
    \abs{K_8} < 47 
    \, ,
\end{align}
where in each case we profiled over the other parameters.
%%%%%%%%%%%%%%%%%%%%%% 
\begin{figure}
\begin{center}
\begin{tabular}{cc}
   \includegraphics[width=0.49\textwidth]{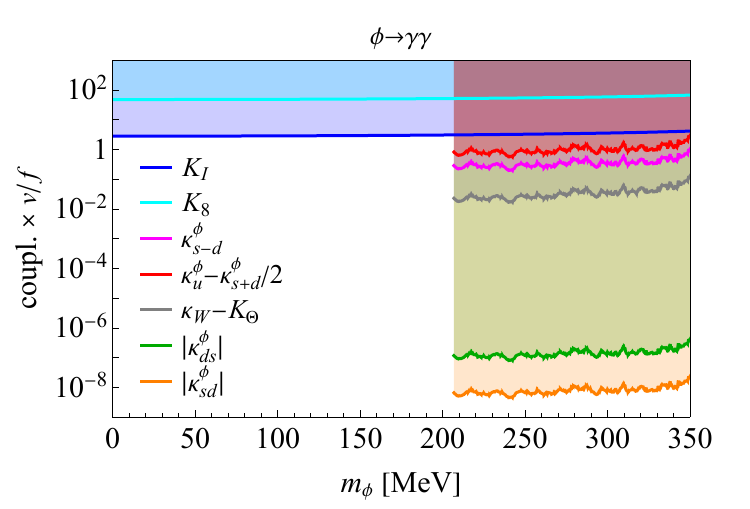}~ & 
    \includegraphics[width=0.49\textwidth]{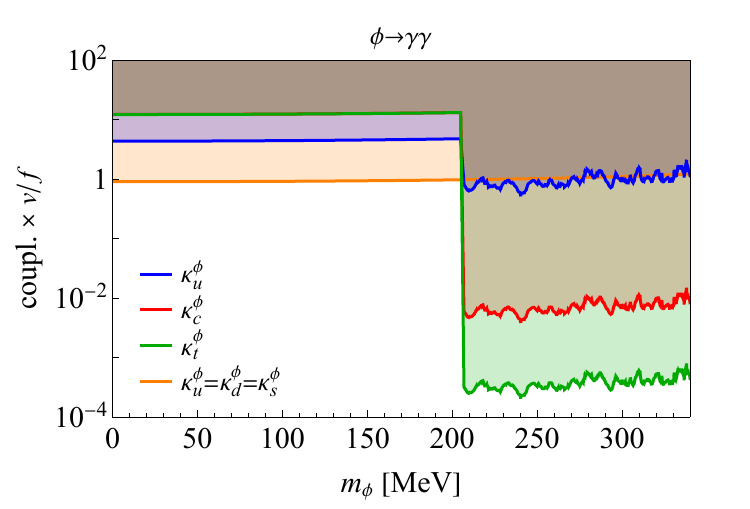}~
\end{tabular}    
    \caption{The 95$\%$ CL upper bounds on $\phi$ couplings to pseudoscalar mesons, assuming  $\phi\to\gamma\gamma$ dominance. (Left) Upper bounds on model-independent $K_I$ and $K_8$ couplings, where the other parameters are profiled over, from $\eta^{(\prime)}$ decays, and 
 on $\kappa_W-K_\Theta$, $\kappa_u^\phi-\kappa^\phi_{s+d}/2$, $\kappa^\phi_{s-d}$ and the penguin contributions $|\kappa^\phi_{sd,ds}|$ from the NA62 search for $K^+\to\pi^+\phi$. 
    Only one parameter is switched on at a time. 
    (Right) Model-dependent upper bounds on $\kappa^\phi_u$, $\kappa^\phi_c$,  $\kappa^\phi_t$ and $\kappa^\phi\equiv\kappa^\phi_u=\kappa^\phi_d=\kappa^\phi_s$ in the $u$--only, $c$--only, $t$--only and $uds$ benchmark models with VLQs, respectively. Note that the limits for the $c,t$--only benchmark models are the same for $m_\phi\lesssim 200\,$MeV.
    }
    \label{fig:KIK8KThetaWUzeta}
\end{center}
\end{figure}
%%%%%%%%%%%%%%%%%%%%%% 
For the constraints from kaon decays, we use the recent NA62 search for $K^+\to\pi^+\phi,\, \phi\to\gamma\gamma$~\cite{NA62:2023olg} to derive bounds on $\kappa_W-K_\Theta$, $\kappa^\phi_{s-d}$, $\kappa^\phi_u+\kappa^\phi_{s+d}/2$ and $|\kappa_{sd,ds}^\phi|$  for $200\,\MeV<m_\phi<350\,\MeV$. Since there are five parameters and only one measurement,  we display in the left panel in \cref{fig:KIK8KThetaWUzeta} the bounds that are obtained when only one parameter at a time is nonzero.  (The bound on ${\rm Im}\,\kappa_{sd}^\phi$  from $K_S\to\pi^0\phi,\, \phi\to\gamma\gamma$ decays is much weaker and is not shown in the figure.)

For the four VLQ model benchmarks introduced in \cref{sec:VLQ},
we set the VLQ mass to $M=5\,\TeV$, in order to evade collider bounds from direct searches for VLQs at the ATLAS and CMS experiments~\cite{ATLAS:2015lpr,CMS:2017asf}. 
For fixed $\kappa^\phi_q$ the remaining dependence on $M$ is only logarithmic. 
The 95$\%$ CL upper bounds on $\kappa_u^\phi$, $\kappa_c^\phi$ and $\kappa_t^\phi$ and $\kappa^\phi_u=\kappa^\phi_d=\kappa^\phi_s\equiv \kappa^\phi$ as a function of $m_\phi$ are given in the right panel of \cref{fig:KIK8KThetaWUzeta}.
For the $uds$ model benchmark, the dominant bound comes from $\eta$ and $\eta'$ decays since in this case the contributions to kaon decays arise only at $\cO(p^4)$ in the weak chiral Lagrangian.
For the other three cases, where VLQ is assumed to either couple only to the up-quark, only the charm-quark or only to the top-quark, the dominant constrains come from $\eta$ and $\eta'$ decays for $m_\phi \lesssim 210\,\MeV$, and from kaon decays for higher $\phi$ masses.

%%%%%%%%%%%%%%%%%%%%%%%%%%%%%%%%%%%%
\subsection{Light scalar decaying invisibly}
\label{sec:phiinv}
%%%%%%%%%%%%%%%%%%%%%%%%%%%%%%%%%%%%

Next, let us assume that $\phi$ decays almost exclusively into dark-sector states or that it is sufficiently long-lived, such that it escapes experimental detection. 
In this case the $\eta$ and $\eta^\prime$ decays do not provide any constraints since $\eta^{(')} \to \pi^0 + \text{invisible}$ decays have not been measured yet.\footnote{Only measured invisible channels are $\eta,\eta' \to \text{invisible}$ and $\eta,\eta' \to \gamma + \text{invisible}$ \cite{Gan:2020aco}.}
The $\phi$ interactions that are contributing to the $K^+\to \pi^+ + \text{invisible}$ channel, on the other hand, are constrained by data from the E949~\cite{BNL-E949:2009dza} and NA62~\cite{NA62:2020pwi,NA62:2020xlg,NA62:2021zjw} experiments for $m_\phi\leq  260\,\MeV$, assuming two-body decay kinematics and efficiencies.
Furthermore, contributions to the $K_L\to \pi^0+ \text{invisible}$ channel are constrained by data from the KOTO experiment~\cite{KOTO:2018dsc,KOTO:2020prk,Koto2023,KOTO:2024zbl}, whose sensitivity can be comparable to that of the NA62 experiment. 

\begin{figure}
\begin{center}
\begin{tabular}{cc}
    \includegraphics[width=0.49\textwidth]{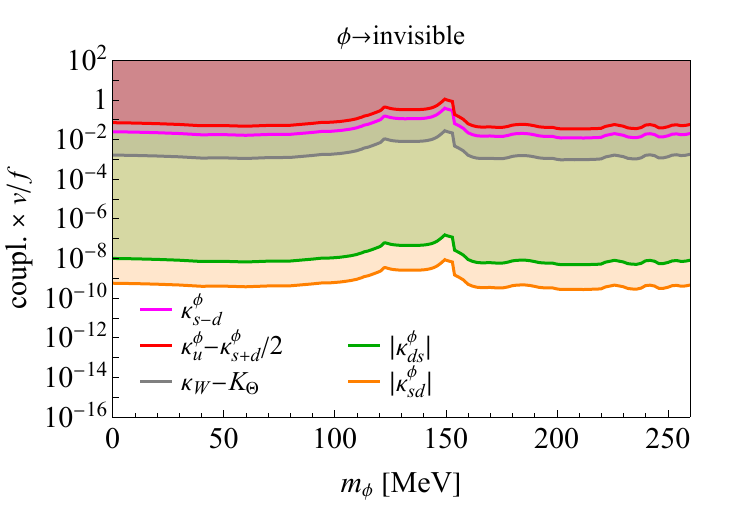}~
    &
    \includegraphics[width=0.49\textwidth]{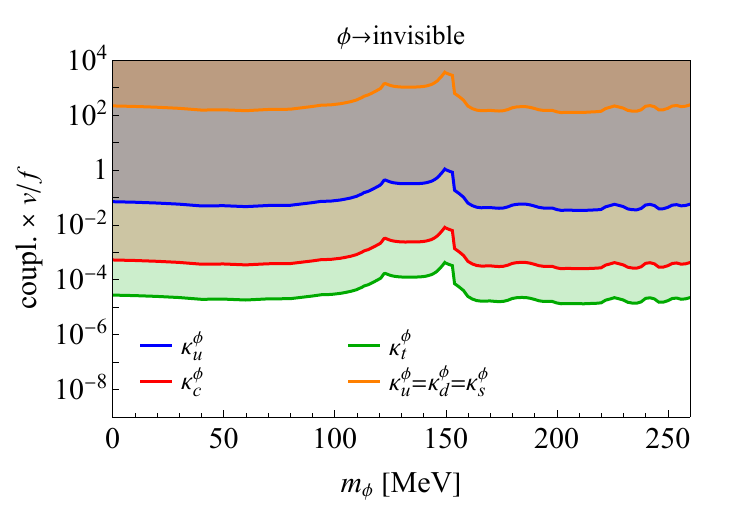}~
\end{tabular}
    \caption{Same as \cref{fig:KIK8KThetaWUzeta}, but assuming $\phi\to\,$invisible dominance.} 
    \label{fig:KIK8KThetaWUzetaINV}
\end{center}
\end{figure}

The 95$\%$ CL upper bounds on $\kappa_W-K_\Theta$, $\kappa_{s-d}^\phi$, $\kappa^\phi_u-\kappa_{s+d}^\phi/2$ and $|\kappa_{sd,ds}^\phi|$, as functions of $m_\phi$ for $m_\phi\leq  260\,\MeV$, are shown in the left panel of \cref{fig:KIK8KThetaWUzetaINV}, taking only a single parameter (out of five) to be nonzero at a time.\footnote{Note that the assumption of invisible $\phi$ is satisfied self-consistently when $\kappa_{sd,ds}^\phi$ are the only nonzero couplings of $\phi$, without the need for $\phi$ to decay invisibly, since it is already long-lived.}
The right panel in \cref{fig:KIK8KThetaWUzetaINV} shows the upper bounds from $K^+\to\pi^+$invisible on the UV parameters, $\kappa_u^\phi$, $\kappa_c^\phi$, $\kappa_t^\phi$ and $\kappa^\phi=\kappa_u^\phi=\kappa_d^\phi=\kappa_s^\phi$ (the $uds$ model) for the four VLQ model benchmarks, setting the VLQ mass to $M=5\,\TeV$.
Note that the bound on $\kappa^\phi$ should be taken only as indicative, since they were estimate using only a partial NLO expression, and thus corrections to up to an order of magnitude are possible. 

%%%%%%%%%%%%%%%%%%%%%% 
%%%%%%%%%%%%%%%%%%%%%% 

%%%%%%%%%%%%%%%%%%%%%%%%%%%%%%%%%%%%
\section{Conclusions}
\label{sec:outro}
%%%%%%%%%%%%%%%%%%%%%%%%%%%%%%%%%%%%

Light scalars, with masses comparable to the QCD scale or well below it, appear in well-motivated extensions of the SM and are a target of many experimental searches.  
However, the current studies mostly focus on two limits: the light Higgs-mixed scalar limit, for the  case of a CP-even spin-$0$ particle, and the ALP limit for the case of a CP-odd light spin-$0$ particle. 
In this manuscript, we introduced the third limit, not covered by the other two: the limit of a \textit{general flavor-aligned light scalar}. 
In the general flavor-aligned light scalar scenario, the couplings of the new particle $\phi$ to the SM fermions do not have any specific structure, except that they are assumed to be aligned with the SM Yukawa couplings. This assumption is introduced purely so that the general flavor-aligned light scalar scenario explores parameter space that is independent of the other two limiting cases, the light Higgs-mixed scalar scenario and the general ALP case. 

The general flavor-aligned light scalar can be realized in concrete UV models. 
In this manuscript we showed this for three examples: in the context of 2HDM in the limit of alignment without decoupling, in a model of a light dark sector dilaton, and on an example of a vector-like quarks with almost flavor aligned structure of couplings and masses. 

Importantly, the phenomenology of the general light flavor aligned scalar differs from the other two commonly considered limits, the light Higgs-mixed scalar and the ALP. 
We limit the discussion to the case of $\phi$ that is not heavier than a few 100\,MeV, so that one can still use ChPT for theoretical predictions, and organize interactions in terms of a chiral expansion. 
Using this effective theory, we calculated the $K\to\pi\phi$, $\eta^{(\prime)}\to\pi\phi$ and $\eta'\to\eta\phi$ decay amplitudes as functions of generic couplings of $\phi$ to the $u$-, $d$-, $s$-quarks and gluons. 
To highlight the phenomenological relevance of our results we considered four benchmark cases for $\phi$ interactions (within a UV complete vector-like quark extension of the SM): the cases where at tree level $\phi$ only couples either to the up quark, only to the charm quark, or only to the top quark, as well as the ``$uds$ model'' benchmark, in which the nonzero couplings of the $\phi$ scalar are assumed to be proportional to the masses of the $u$, $d$ and $s$ quarks, respectively.  The numerical results for these benchmark are summarized in \cref{sec:bounds}. 
The limit of $\phi$ coupling only to $uds$ is quite distinct;
the $K^+\to\pi^+\phi$ decays arise only at NLO in chiral expansion, and thus the constraints from $K^+\to\pi^++$invisible and from $K^+\to\pi^+\gamma\gamma$ can largely be avoided. 
In this limit other searches, not discussed in detail in this manuscript, such as the precision measurement of atomic and molecular systems, can become important, see, \eg, Ref.~\cite{Delaunay:2022grr}. 

There are several directions in which our initial analysis of the general flavor aligned light scalar could be improved upon.  
For heavier $\phi$ the decays of heavier mesons, such as $D\to \pi \phi$, $D\to K\phi$, $B\to K\phi$, etc, would become important. 
Furthermore, for heavier $\phi$ the other decays beyond $\phi\to \gamma\gamma$ and $\phi\to$invisible, already considered here, could become dominant, depending on the assumed flavor diagonal couplings. Finally, it would be interesting to revisit astrophysical and cosmological bounds in this more general context. 

%%%%%%%%%%%%%%%%%%%%%%%%%%%%%%%%%%%%
\section*{Acknowledgments}
%%%%%%%%%%%%%%%%%%%%%%%%%%%%%%%%%%%%

We are indebted to Diego Redigolo for collaboration during initial stages of this work, and to Brian Batell for sharing the details about their results in \cite{Batell:2017kty}. 
We thank Yael Shadmi for enlightening discussion about VLQ models.  
The work of C.\,D. is supported by the CNRS IRP NewSpec. 
T.\,K. is supported by the Grant-in-Aid for Scientific Research from the Ministry of Education, Culture, Sports, Science, and Technology (MEXT), Japan, No.\,21K03572 and No.\,24K22872.
The work of T.\,K. is also supported by the Japan Society for the Promotion of Science (JSPS)  Core-to-Core Program, No.\,JPJSCCA20200002.  
J.Z. acknowledges support in part by NSF grants OAC-2103889, OAC-2411215, OAC-2417682, and DOE grant DE-SC101977.
The work of Y.S. is supported by grants from the NSF-BSF (grant No. 2021800), the ISF (grant No. 597/24).

%%%%%%%%%%%%%%%%%%%%%%%%%%%%%%%%%%%%
%%%%%%%%%%%%%%%%%%%%%%%%%%%%%%%%%%%%
\appendix
%%%%%%%%%%%%%%%%%%%%%%%%%%%%%%%%%%%%
%%%%%%%%%%%%%%%%%%%%%%%%%%%%%%%%%%%%

%%%%%%%%%%%%%%%%%%%%%%%%%%%%%%%%%%%%
\section{Neutral meson mass eigenstates}
\label{sec:pietamixing}
%%%%%%%%%%%%%%%%%%%%%%%%%%%%%%%%%%%%

In this appendix we perform a mass diagonalization for the neutral mesons. Let us denote the interaction states as $\hat P_i=(\pi^0_{\rm int},\eta_8, \eta_0$), and the mass eigenstates as $P_i=(\pi^0,\eta,\eta'$) (note that in \cref{eq:Pi:explicit} in the main text $\pi^0_{\rm int}$ was denoted simply as $\pi^0$ for brevity). In the interaction basis, the mass matrix for the neutral mesons is at the leading order given by 
\begin{align}
    M_{P}^2
    =
    \begin{pmatrix} 
        2B_0\hat m& 
        -\frac{2}{\sqrt{3}}B_0\hat m \delta_I & 
        -\frac{2\sqrt{2}}{\sqrt{3}}B_0\hat m \delta_I \\ 
        -\frac{2 }{\sqrt{3}}B_0\hat m\delta_I& 
        \frac{2}{3}B_0(\hat m +2m_s) & 
        \frac{2\sqrt{2}}{3}B_0(\hat m -m_s)\\
        -\frac{2\sqrt{2}}{\sqrt{3}}B_0\hat m \delta_I & 
        \frac{2\sqrt{2}}{3}B_0(\hat m -m_s) & 
        \frac{2}{3}B_0(2\hat m +m_s)+\mu_{\eta_0}^2
    \end{pmatrix}\,,
\end{align}
where $\hat m \equiv (m_u+m_d)/2$ and $\delta_I\equiv (m_d-m_u)/(m_u+m_d)$ are the average and the relative difference of the up and down quark masses, respectively. 

In the isospin limit, $\delta_I\to 0$, the $\pi_{\rm int}^0$ does not mix with the $\eta_{8,0}$ mesons ($m_{\pi^0}^2=2B_0\hat m$) and the $\eta$ and $\eta'$ mass eigenstates are
\begin{align}
    \begin{pmatrix} \eta \\ \eta' \end{pmatrix} 
    = 
    \begin{pmatrix} 
        \cos\theta_{\eta\eta'} & -\sin\theta_{\eta\eta'} \\ 
        \sin\theta_{\eta\eta'} & \cos\theta_{\eta\eta'}
    \end{pmatrix}
    \begin{pmatrix} \eta_8 \\ \eta_0  \end{pmatrix}  \, ,
\end{align}
where $\sin2\theta_{\eta\eta'}<0$ and
\begin{align}
    \tan2\theta_{\eta\eta'}
    =   
    \frac{4\sqrt{2}B_0(\hat m -m_s)}{2B_0(\hat m-m_s)+3 \mu_{\eta_0}^2}\,,
\end{align}
with masses
\begin{align}
    m_{\eta,\eta'}^2
    = 
    B_0(\hat m+m_s)+\frac{1}{2}\mu_{\eta_0}^2 \mp\frac{2\sqrt{2}B_0(\hat m-m_s)}{3\sin2\theta_{\eta\eta'}}\,.
\end{align}
Isospin violation ($\delta_I\neq 0$) further induces mass mixings between the $\eta,\eta'$ mesons and the $\pi^0_{\rm int}$. 
To linear order in $\delta_I$, the above expressions for the meson masses remain unchanged, while the mass and the interaction  states are now related as
\begin{align}
\label{eq:pi0:expr}
    \pi_{\rm int}^0
    &=  
    {\pi}^0- (\vartheta_{\pi\eta} \eta +\vartheta_{\pi\eta'} \eta')\,,\\ 
     \label{eq:eta8:expr}
    \eta_8
    &=  
    \cos\theta_{\eta\eta'}\eta+ \sin\theta_{\eta\eta'}\eta'+(\vartheta_{\pi\eta}\cos\theta_{\eta\eta'}+\vartheta_{\pi\eta'}\sin\theta_{\eta\eta'})\pi^0\,,\\
    \label{eq:eta0:expr}
    \eta_0
    &=  
    \cos\theta_{\eta\eta'}\eta'-\sin\theta_{\eta\eta'}\eta+ (\vartheta_{\pi\eta'}\cos\theta_{\eta\eta'}-\vartheta_{\pi\eta}\sin\theta_{\eta\eta'})\pi^0 \,,
\end{align}
where
\begin{align}
    \vartheta_{\pi\eta}
    =
    \delta_I\frac{m_\pi^2\cos(\theta_{\eta\eta'}+\alpha)}{m_\eta^2-m_\pi^2}\,,
    \qquad 
    \vartheta_{\pi\eta'}
    =
    \delta_I\frac{m_\pi^2\sin(\theta_{\eta\eta'}+\alpha)}{m_{\eta'}^2-m_\pi^2}\,,
\end{align}
and $\alpha\equiv \arctan\sqrt{2}\simeq 0.9553$.

Note that at leading order in ChPT there aren't enough free parameters in the chiral Lagrangian to simultaneously accommodate both the experimental value of the $\eta'$ mass as well as the $\eta/\eta'$ mixing angle~\cite{Georgi:1993jn}. 
Reproducing the latter requires inclusion of additional operators that are of higher-order in the chiral expansion~\cite{Peris:1993np,Gerard:2004gx}. 
In this manuscript we use in the calculations of the transition matrix element the leading order relations between the interaction and mass eigenstates, \cref{eq:pi0:expr,eq:eta8:expr,eq:eta0:expr}, setting the $\eta,\eta'$ masses and the $\theta_{\eta\eta'}$ mixing angle to their measured values: $m_\eta\simeq 548\,\MeV$, $m_{\eta'}\simeq 958\,\MeV$~\cite{ParticleDataGroup:2024cfk} and $\theta_{\eta\eta'}\simeq-(22\pm1)^{\circ}$~\cite{Gerard:2005yk}. 
In particular, the mass diagonalization corrects the $K_{L,S}\to \pi^0\phi$ decay amplitudes by a common factor $1-R_{\pi^0-\eta/\eta'}$, see \cref{eq:pi-eta-eta':mixing}, where to leading order in $\delta_I$, 
\begin{align}
    R_{\pi^0-\eta/\eta'}
    =& \frac{1}{\sqrt{3}}\left[(2\sqrt{2}\vartheta_{\pi\eta'}-\vartheta_{\pi\eta})\cos\theta_{\eta\eta'}-(2\sqrt{2}\vartheta_{\pi\eta}+\vartheta_{\pi\eta'})\sin\theta_{\eta\eta'}\right]\nonumber\\
=&    \delta_I\frac{m_{\pi^0}^2\left[m_{\eta'}^2+m_\eta^2-2m_{\pi^0}^2+\sqrt{3}(m_{\eta'}^2-m_\eta^2)\cos(2\theta_{\eta\eta'}+3\alpha)\right]}{2(m_\eta^2-m_{\pi^0}^2)(m_{\eta'}^2-m_{\pi^0}^2)}
    \approx 8.6\times 10^{-3}\,.
\end{align}
%

%%%%%%%%%%%%%%%%%%%%%%%%%%%%%%%%%%%%
\section{Derivation of ChPT Lagrangian for $\phi$ interactions}
\label{sec:ChPT:derivation}
%%%%%%%%%%%%%%%%%%%%%%%%%%%%%%%%%%%%

In this appendix we give details about the derivation of the ChPT Lagrangian for the interactions of the light scalar $\phi$ in \cref{sec:CPevenInt}, \cref{eq:Leffphi}. 
We follow the approach of Ref.~\cite{Leutwyler:1989xj}, where the scalar currents were first rewritten in terms of the trace of the energy-momentum tensor and the remainder, and then each of the contributions was hadronized using ChPT rules. 

The trace of the energy-momentum tensor, $\Theta_\mu^{\ \mu}$, is related to the divergence of a dilatation current and vanishes in scale-invariant theories~\cite{Callan:1970ze}. 
At $\cO(\GeV)$ scale invariance is broken by the running of the QCD and QED gauge couplings, the finite quark and lepton masses, and by the $\Delta S=1$ four-quark operators, $\cL_{4q}^{\Delta S=1}$, cf. \cref{eq:DeltaS=1}. 
This gives~\cite{Collins:1976yq}
\begin{align}
    \label{eq:ThetaQCD}
    \Theta_\mu^{\ \mu} 
    = 
    -b\frac{\alpha_s}{8\pi}G_{\mu\nu}^aG^{\mu\nu\,a}
    -\tilde{b}\frac{\alpha}{8\pi}F_{\mu\nu}F^{\mu\nu}
    +\sum_{\psi=q,\ell} m_\psi\overline \psi \psi
    +2\cL_{4q}^{\Delta S=1}\,,
\end{align}  
where $b=11-2n_q/3$ and $\tilde b=-4(\sum_q  Q_q^2+n_\ell/3)$ are the leading order~(LO) QCD and QED $\beta$-function coefficients, respectively. 
In the following, we take $n_q=3$ (for $u,d,s$ quarks) and $n_\ell=2$ (for electron and muon), so that $b=9$ and $\tilde{b}=-16/3$.  
In \eqref{eq:ThetaQCD} we also keep only the LO anomalous dimension for the running of the fermion mass terms, and similarly, at LO the coefficient of the $\cL_{4q}^{\Delta S=1}$ in \eqref{eq:ThetaQCD} is simply given by the scaling dimension of the four-quark operators. 
In writing down $\Theta_\mu^\mu$ we can neglect the $\phi$ interactions \cref{eq:Lint2GeV} and \eqref{eq:Q12L}, since these correspond to nonlinear $\phi$ interactions in the chiral Lagrangian, which are of higher order in $1/f$ counting. 

Using \cref{eq:ThetaQCD} we can rewrite the WET${}_\phi$ interactions of the $\phi$ scalar, \cref{eq:Lint2GeV,eq:Q12L}, as  
\beq
\begin{split}
    \label{eq:Lint1GeVK:app}
    \cL_{\rm int}^\phi= \cL_{\rm int}^{\rm diag}+ \cL_{\rm int}^{sd}
    =&
    -\frac{\phi}{f}\biggr[
    K_\Theta \Theta_\mu^{\ \mu}
    +K_\gamma F_{\mu\nu}F^{\mu\nu}
    +\sum_{\psi=q,\ell} K_\psi m_\psi \bar \psi \psi 
    \\
    &   \qquad
    -\big(\kappa_{sd}^\phi m_s\bar d_L s_R+\kappa_{ds}^\phi m_d\bar d_R s_L+{\rm h.c.} \big) +K_W \cL_{4q}^{\Delta S=1}
    \biggr],
\end{split}
\eeq
where the sum over $\psi$ is over the light quark flavors, $q=u,d,s$, and the two lightest charged leptons $\ell=e, \mu$, and
\begin{align}
    K_\Theta&=\frac{8\pi}{\alpha_s b} c_G^\phi=\frac{8\pi}{9 \alpha_s} c_G^\phi\, , \\
    K_\gamma&=-c_\gamma^\phi +\tilde b \frac{\alpha}{8\pi } K_\Theta=-c_\gamma^\phi -\frac{2\alpha}{3\pi } K_\Theta\, , \\
    K_\psi &= \kappa_\psi^\phi- K_\Theta\, , \\
    K_W&=2( \kappa_W-K_\Theta) \, .
\end{align}

Following Ref.~\cite{Leutwyler:1989xj} we can now hadronize separately each of the operators in \cref{eq:Lint1GeVK:app}, treating $\phi$ as an external classical field, which is an approximation valid in the limit of a feebly interacting light $\phi$ field, that we are interested in. 
We start with the trace of the energy-momentum tensor.
In ChPT, supplemented by the two lightest leptons, and with QED interactions included, this is given by 
\begin{align}
    \label{eq:Thetaconfined}
    -\frac{\phi}{f}
    K_\Theta  \Theta_\mu^{\ \mu}
    \to 
   -\frac{\phi}{f}
    K_\Theta\biggr( \bar\Theta_\mu^{\ \mu}
    -\tilde b_{\rm eff} \frac{\alpha}{8\pi}F_{\mu\nu}F^{\mu\nu}
    +\sum_{\ell}m_\ell \bar\ell \ell\biggr), 
\end{align}
where the arrow denotes hadronization, i.e., on the l.h.s. we have a single insertion of the $\phi$ interaction, evaluated for any transitions with arbitrary insertions of the QCD Lagrangian, while on the r.h.s. the same transition amplitude is to be evaluated in ChPT at ${\mathcal O}(p^2)$. Above, $\tilde b_{\rm eff}=-(n_M+4n_\ell)/3=-10/3$ is the QED $\beta$-function for this low energy EFT, with $n_M=2$ the number of charged light mesons ($\pi^\pm$ and $K^\pm$), and $n_\ell=2$ the number of light charged leptons. 
In \cref{eq:Thetaconfined} $\bar \Theta_\mu^\mu$ is the trace of the energy-momentum tensor in ChPT, which is given by~\cite{Donoghue:1991qv} 
\begin{align}
    \label{eq:ThetaXPT}
    \bar\Theta_{\mu\nu}
    = 
    \frac{f_\pi^2}{8}\big\langle{\left(1 - \gamma_1 \lambda_6\right)
    \big(D_\mu U^\dagger D_\nu U
    + D_\nu U^\dagger D_\mu U\big)}\big\rangle-g_{\mu\nu}\cL_{\rm eff}(U)\,,
\end{align}
where $\gamma_1$ is the low energy Wilson coefficient in the weak ChPT Lagrangian, \cref{eq:Leffweak}, $\lambda_6$ is the Gell-Mann SU(3) matrix, and $\cL_{\rm eff}(U)= \cL_{\rm eff}^{\rm str.}(U)+\cL_{\rm eff}^{\Delta S=1}(U)$. The ChPT expression for the first term in \cref{eq:Lint1GeVK:app} is thus given by $-\frac{\phi}{f} K_\Theta \Theta_\mu^{\ \mu}$, where $\Theta_\mu^{\ \mu}$ is given in \cref{eq:Thetaconfined}, with 
\beq
\bar\Theta_\mu^{\ \mu}
    = 
    \frac{f_\pi^2}{4}\big\langle\left(1 - \gamma_1 \lambda_6\right)
    D_\mu U^\dagger D^\mu U \big\rangle-4\cL_{\rm eff}(U)\,.
\eeq

The interactions of $\phi$ with the quarks can be written using a $\phi$ dependent $3\times3$ mass matrix as, 
\begin{align}
   - \frac{\phi}{f}\Big[\sum_{q=u,d,s} K_q m_q  \bar q q 
    -\big(\kappa_{sd}^\phi m_s\bar d_L s_R+\kappa_{ds}^\phi m_d\bar d_R s_L+{\rm h.c.} \big)\Big] 
    =
    -\frac{\phi}{f} \bar q_R M_\kappa q_L +{\rm h.c.} \, ,
\end{align}
where $q=(u,d,s)$, and (see also \cref{eq:cMK:longer})
\begin{align}
\label{eq:cMK}
    M_\kappa=
    \begin{pmatrix}
        K_u m_u & 0 & 0  \\
        0 & K_d m_d &-\kappa_{ds}^\phi m_d \\
        0& -\kappa_{sd}^{\phi *} m_s & K_s m_s
    \end{pmatrix} \, .
\end{align}
Using the same spurion technique as for $\cL_{\rm eff}^{\rm str.}(U)$, \cref{eq:Leff}, and $\cL_{\rm eff}^{\Delta S=1}(U)$ in \cref{eq:Leffweak}, we can now replace the operators describing scalar interactions of quarks with $\phi$ in \cref{eq:Lint1GeVK:app}  with the hadronized ChPT version of these operators,
\beq
    \begin{split}
    -\bar q_R \frac{\phi}{f} M_\kappa q_L(0) +{\rm h.c.} &-\int d^4 x T\{\bar q_R \frac{\phi}{f} M_\kappa q_L(0)+{\rm h.c.}, i\cL_{4q}^{\Delta S=1}(x)\}  
    \to 
    \\
   & \frac{f_\pi^2}{4}B_0 \frac{\phi}{f} \big\langle{ \left( 1 - \gamma_2 \lambda_6 \right) \left( M_\kappa^\dagger U 
    + U^\dagger M_\kappa\right) }\big\rangle,
    \end{split}
\eeq
where on the l.h.s. we work to linear order in the $\phi$ interactions, and also to  first order in the weak interaction insertions.
Finally, the weak interaction term $\cL_{4q}^{\Delta S=1}$ in \cref{eq:Lint1GeVK:app} can be replaced by its ChPT version, $\cL_{4q}^{\Delta S=1}\to \cL_{\rm eff}^{\Delta S=1}(U)$, with $\cL_{\rm eff}^{\Delta S=1}(U)$ given in \cref{eq:Leffweak}. 
Collecting all the above terms gives \cref{eq:Leffphi} in the main text. 

%%%%%%%%%%%%%%%%%%%%%%%%%%%%%%%%%%%%
\section{Feynman rules for the VLQ${}_\phi$ model}
\label{sec:Feynman-rules}
%%%%%%%%%%%%%%%%%%%%%%%%%%%%%%%%%%%%

In this appendix, we list the Feynman rules for the renormalizable VLO${}_\phi$ model, where one adds to the SM three generations of up- and down-type VLQ isospin singlets and a light scalar singlet, $\phi$. The Lagrangian of the model is given in \cref{eq:LVLQ}.

Table~\ref{tab:feynman-rules} gathers the charged-current interactions, coupling $\phi$ to the $W^\pm$ gauge fields and to the $G^\pm$  Goldstone fields, while \cref{tab:feynman-rules2} gives the Feynman rules for $\phi$ interactions with SM fermions and VLQs, all expressed in the fermion mass basis. The dominant terms at leading order in the SM quark/VLQ mixing angles are also given.

\begin{center}
\begin{table}
\centering
\begin{tabular}{cl}
\hline\\
\scalebox{0.8}{\begin{tikzpicture}[baseline=(b1.base)]
  \begin{feynman}
    \vertex (b1) {$W^-_\mu$};
    \vertex [right= of b1] (b2);
    \vertex [right=2em of b2] (b5);
    \vertex [above=3em of b5] (b3) {$u_i$};
    \vertex [below=3em of b5] (b4) {$d_j$};
    
\diagram*{
(b1) -- [charged boson] (b2),
(b3) -- [fermion, arrow size=1pt] (b2) -- [fermion, arrow size=1pt] (b4)
};
\end{feynman}
\end{tikzpicture}
}
&
$\begin{aligned}
&=-i\frac{g}{\sqrt{2}}\gamma_\mu P_L\cos\theta_{u_L^i}\cos\theta_{d_L^j}V_{ij}^*\\
&\simeq -i\frac{g}{\sqrt{2}}\gamma_\mu P_L V_{ij}^*
\end{aligned}
$\\

\scalebox{0.8}{\begin{tikzpicture}[baseline=(b1.base)]
  \begin{feynman}
    \vertex (b1) {$W^-_\mu$};
    \vertex [right= of b1] (b2);
    \vertex [right=2em of b2] (b5);
    \vertex [above=3em of b5] (b3) {$U_i$};
    \vertex [below=3em of b5] (b4) {$d_j$};
    
\diagram*{
(b1) -- [charged boson] (b2),
(b3) -- [fermion, very thick, arrow size=1pt] (b2) -- [fermion, arrow size=1pt] (b4)
};
\end{feynman}
\end{tikzpicture}
}
&
$\begin{aligned}
&=-i\frac{g}{\sqrt{2}}\gamma_\mu P_L\sin\theta_{u_L^i}\cos\theta_{d_L^j}V_{ij}^*\\
&\simeq -i\frac{g}{\sqrt{2}}\gamma_\mu P_L \theta_{u_L^i} V_{ij}^* 
\end{aligned}$\\

\scalebox{0.8}{\begin{tikzpicture}[baseline=(b1.base)]
  \begin{feynman}
    \vertex (b1) {$W^-_\mu$};
    \vertex [right= of b1] (b2);
    \vertex [right=2em of b2] (b5);
    \vertex [above=3em of b5] (b3) {$u_i$};
    \vertex [below=3em of b5] (b4) {$D_j$};
    
\diagram*{
(b1) -- [charged boson] (b2),
(b3) -- [fermion, arrow size=1pt] (b2) -- [fermion, very thick, arrow size=1pt] (b4)
};
\end{feynman}
\end{tikzpicture}
}
&
$\begin{aligned}
&=-i\frac{g}{\sqrt{2}}\gamma_\mu P_L\cos\theta_{u_L^i}\sin\theta_{d_L^j}V_{ij}^*\\
&\simeq -i\frac{g}{\sqrt{2}}\gamma_\mu P_L \theta_{d_L^j} V_{ij}^* 
\end{aligned}$\\

\scalebox{0.8}{\begin{tikzpicture}[baseline=(b1.base)]
  \begin{feynman}
    \vertex (b1) {$W^-_\mu$};
    \vertex [right= of b1] (b2);
    \vertex [right=2em of b2] (b5);
    \vertex [above=3em of b5] (b3) {$U_i$};
    \vertex [below=3em of b5] (b4) {$D_j$};
    
\diagram*{
(b1) -- [charged boson] (b2),
(b3) -- [fermion, very thick, arrow size=1pt] (b2) -- [fermion, very thick, arrow size=1pt] (b4)
};
\end{feynman}
\end{tikzpicture}
}
&
$\begin{aligned}
&=-i\frac{g}{\sqrt{2}}\gamma_\mu P_L\sin\theta_{u_L^i}\sin\theta_{d_L^j}V_{ij}^*\\
&\simeq \cO(\theta^2)
\end{aligned}$\\

\scalebox{0.8}{\begin{tikzpicture}[baseline=(b1.base)]
  \begin{feynman}
    \vertex (b1) {$G^-$};
    \vertex [right= of b1] (b2);
    \vertex [right=2em of b2] (b5);
    \vertex [above=3em of b5] (b3) {$u_i$};
    \vertex [below=3em of b5] (b4) {$d_j$};
    
\diagram*{
(b1) -- [charged scalar] (b2),
(b3) -- [fermion, arrow size=1pt] (b2) -- [fermion, arrow size=1pt] (b4)
};
\end{feynman}
\end{tikzpicture}
}
&
$\begin{aligned}
&=    i\left[\cos\theta_{d_L^j}\left(y_u^i\cos\theta_{u_R^i}-y_U^i\sin\theta_{u_R^i}\right)P_R\right.\\
&\quad \left.-\cos\theta_{u_L^i}\left(y_d^j \cos\theta_{d_R^j}-y_D^j\sin\theta_{d_L^j}\right)P_L\right]V_{ij}^*\\
& \simeq i\frac{\sqrt{2}}{v}\left(m_{u_i}P_R-m_{d_j}P_L\right)V_{ij}^*
\end{aligned}
$\\

\scalebox{0.8}{\begin{tikzpicture}[baseline=(b1.base)]
  \begin{feynman}
    \vertex (b1) {$G^-$};
    \vertex [right= of b1] (b2);
    \vertex [right=2em of b2] (b5);
    \vertex [above=3em of b5] (b3) {$U_i$};
    \vertex [below=3em of b5] (b4) {$d_j$};
    
\diagram*{
(b1) -- [charged scalar] (b2),
(b3) -- [fermion, very thick, arrow size=1pt] (b2) -- [fermion, arrow size=1pt] (b4)
};
\end{feynman}
\end{tikzpicture}
}
&
$\begin{aligned}
&=    i\left[\cos\theta_{d_L^j}\left(y_u^i\sin\theta_{u_R^i}+y_U^i\cos\theta_{u_R^i}\right)P_R\right.\\
&\quad \left.-\sin\theta_{u_L^i}\left(y_d^j \cos\theta_{d_R^j}-y_D^j\sin\theta_{d_L^j}\right)P_L\right]V_{ij}^*\\
&\simeq i\frac{\sqrt{2}}{v}\left(m_{U_i}P_R-m_{d_j}P_L\right)\theta_{u_L^i}V_{ij}^*
\end{aligned}$
\\

\scalebox{0.8}{\begin{tikzpicture}[baseline=(b1.base)]
  \begin{feynman}
    \vertex (b1) {$G^-$};
    \vertex [right= of b1] (b2);
    \vertex [right=2em of b2] (b5);
    \vertex [above=3em of b5] (b3) {$u_i$};
    \vertex [below=3em of b5] (b4) {$D_j$};
    
\diagram*{
(b1) -- [charged scalar] (b2),
(b3) -- [fermion, arrow size=1pt] (b2) -- [fermion, very thick, arrow size=1pt] (b4)
};
\end{feynman}
\end{tikzpicture}
}
&
$\begin{aligned}
&=    i\left[\sin\theta_{d_L^j}\left(y_u^i\cos\theta_{u_R^i}-y_U^i\sin\theta_{u_R^i}\right)P_R\right.\\
&\quad\left.-\cos\theta_{u_L^i}\left(y_d^j \sin\theta_{d_R^j}+y_D^j\cos\theta_{d_L^j}\right)P_L\right]V_{ij}^*\\
& \simeq i\frac{\sqrt{2}}{v}\left(m_{u_i}P_R-m_{D_j}P_L\right)\theta_{d_L^j}V_{ij}^*
\end{aligned}$
\\
\hline
\end{tabular}
\caption{Feynman rules for the charged currents in the VLQ${}_\phi$ model. The chirality projectors are $P_{L,R}\equiv (1\mp \gamma_5)/2$. For the incoming $W^+$ line the Feynman rule follows from the one for $W^-$ by making the $V_{ij}^*\to V_{ij}$ replacement, while the Feynman rules for the incoming $G^+$ follow from the one for $G^-$ by replacing $P_L\leftrightarrow P_R$. }
\label{tab:feynman-rules}
\end{table}
\end{center}

\begin{center}
\begin{table}
\centering
\begin{tabular}{cl}
\hline\\

\scalebox{0.8}{\begin{tikzpicture}[baseline=(b1.base)]
  \begin{feynman}
    \vertex (b1) {$\phi$};
    \vertex [right= of b1] (b2);
    \vertex [right=2em of b2] (b5);
    \vertex [above=3em of b5] (b3) {$f_i$};
    \vertex [below=3em of b5] (b4) {$f_j$};
    
\diagram*{
(b1) -- [scalar] (b2),
(b3) -- [fermion, arrow size=1pt] (b2) -- [fermion, arrow size=1pt] (b4)
};
\end{feynman}
\end{tikzpicture}
}
&
$\begin{aligned}
&=i\lambda_F^i\cos\theta_{f_R^i}\sin\theta_{f_L^i}\delta_{ij}\\
&\simeq i\lambda_F^i\theta_{f_L^i}\delta_{ij}
\end{aligned}$\\

\scalebox{0.8}{\begin{tikzpicture}[baseline=(b1.base)]
  \begin{feynman}
    \vertex (b1) {$\phi$};
    \vertex [right= of b1] (b2);
    \vertex [right=2em of b2] (b5);
    \vertex [above=3em of b5] (b3) {$F_i$};
    \vertex [below=3em of b5] (b4) {$f_j$};
    
\diagram*{
(b1) -- [scalar] (b2),
(b3) -- [fermion, very thick, arrow size=1pt] (b2) -- [fermion, arrow size=1pt] (b4)
};
\end{feynman}
\end{tikzpicture}
}
&
$\begin{aligned}
&=-i\lambda_F^i\left(\cos\theta_{f_L^i}\cos\theta_{f_R^i}P_L-\sin\theta_{f_L^i}\sin\theta_{f_R^i}P_R\right)\delta_{ij}\\
&\simeq -i\lambda_F^iP_L\delta_{ij}
\end{aligned}$\\

\scalebox{0.8}{\begin{tikzpicture}[baseline=(b1.base)]
  \begin{feynman}
    \vertex (b1) {$\phi$};
    \vertex [right= of b1] (b2);
    \vertex [right=2em of b2] (b5);
    \vertex [above=3em of b5] (b3) {$f_i$};
    \vertex [below=3em of b5] (b4) {$F_j$};
    
\diagram*{
(b1) -- [scalar] (b2),
(b3) -- [fermion, arrow size=1pt] (b2) -- [fermion, very thick, arrow size=1pt] (b4)
};
\end{feynman}
\end{tikzpicture}
}
&
$\begin{aligned}
&=-i\lambda_F^i\left(\cos\theta_{f_L^i}\cos\theta_{f_R^i}P_R-\sin\theta_{f_L^i}\sin\theta_{f_R^i}P_L\right)\delta_{ij}\\
&\simeq -i\lambda_F^i P_R\delta_{ij}
\end{aligned}$\\

\scalebox{0.8}{\begin{tikzpicture}[baseline=(b1.base)]
  \begin{feynman}
    \vertex (b1) {$\phi$};
    \vertex [right= of b1] (b2);
    \vertex [right=2em of b2] (b5);
    \vertex [above=3em of b5] (b3) {$F_i$};
    \vertex [below=3em of b5] (b4) {$F_j$};
    
\diagram*{
(b1) -- [scalar] (b2),
(b3) -- [fermion, very thick, arrow size=1pt] (b2) -- [fermion, very thick, arrow size=1pt] (b4)
};
\end{feynman}
\end{tikzpicture}
}
&
$\begin{aligned}
&=i\lambda_F^i\sin\theta_{f_R^i}\cos\theta_{f_L^i}\delta_{ij}\\
&\simeq i\lambda_F^i\frac{m_{f_i}}{m_{F_i}}\theta_{f_L^i}\delta_{ij}
\end{aligned}$
\\
\hline
\end{tabular}
\caption{Feynman rules for $\phi$ interactions in the VLQ${}_\phi$ model. The $f,F$ states denotes both $u,U$ and $d,D$.
}
\label{tab:feynman-rules2}
\end{table}
\end{center}

%%%%%%%%%%%%%%%%%%%%%%%%%%%%%%%%%%%%
\section{One-loop penguins for down-type VLQs}
\label{sec:renormalization}
%%%%%%%%%%%%%%%%%%%%%%%%%%%%%%%%%%%%

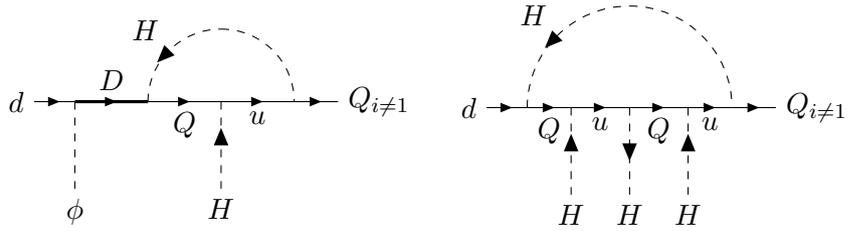
\begin{figure}
\begin{center}
\begin{tabular}{cc}

\begin{tikzpicture}
  \begin{feynman}
    \vertex (b1) {$d$};
    \vertex [right=2em of b1] (b2);
    \vertex [right=2.5em of b2] (b3);
    \vertex [right=2.5em of b3] (b4);
    \vertex [right=2.5em of b4] (b5);
    \vertex [right=1.5em of b5] (b6) {$Q_{i\neq 1}$};
    \vertex[below=3em of b2] (s1) {$\phi$};
\vertex[below=3em of b4] (t1) {$H$};
    \vertex[above=2.5em of b3](g2);
\vertex[right=2.5em of g2](g1);

\diagram* {
(b1)  -- [fermion,  arrow size=1pt] (b2)  -- [fermion, edge label={$D$}, very thick, arrow size=1pt ] (b3) -- [fermion, edge label'={$Q$}, arrow size=1pt] (b4) -- [fermion, edge label'={$u$}, arrow size=1pt] (b5) -- [fermion, arrow size=1pt] (b6),
(b3) -- [anti charged scalar, quarter left, edge label={\(H\)}] (g1) -- [scalar, quarter left] (b5), 
(b2) -- [scalar] (s1),
(b4) -- [anti charged scalar] (t1)
};
 
\end{feynman}
\end{tikzpicture}

&

\begin{tikzpicture}
  \begin{feynman}
    \vertex (b1) {$d$};
    \vertex [right=2em of b1] (b2);
    \vertex [right=1.5em of b2] (b3);
    \vertex [right=2em of b3] (b4);
    \vertex [right=2em of b4] (b5);
    \vertex [right=1.5em of b5] (b6);
    \vertex [right=1.5em of b6] (b7) {$Q_{i\neq 1}$};
    \vertex[below=3em of b3] (s1) {$H$};
    \vertex[below=3em of b4] (t1) {$H$};
    \vertex[below=3em of b5] (x1) {$H$};
    \vertex[above=3.5em of b3](g2);
\vertex[right=2em of g2](g1);

\diagram* {
(b1)  -- [fermion,  arrow size=1pt] (b2)  -- [fermion, arrow size=1pt, edge label'={$Q$} ] (b3) -- [fermion, edge label'={$u$}, arrow size=1pt] (b4) -- [fermion, edge label'={$Q$}, arrow size=1pt] (b5) -- [fermion, arrow size=1pt, edge label'={$u$}] (b6) -- 
 [fermion, arrow size=1pt] (b7),
(b2) -- [anti charged scalar, quarter left, edge label={\(H\)}] (g1) -- [scalar, quarter left] (b6), 
(b3) -- [anti charged scalar] (s1),
(b4) -- [charged scalar] (t1),
(b5) -- [anti charged scalar] (x1)
};

\end{feynman}
\end{tikzpicture}

\end{tabular}
\caption{ One-loop diagrams contributing to scalar FCNC in the down-type VLQ model of Section~\ref{sec:UV} (left) and the SM (right). Arrows indicate the hypercharge flow.}
\label{fig:UVpenguin}
\end{center}
\end{figure}

The assumed flavor structure for VLQ model in \cref{eq:yUhat} does not suffice to renormalize the VLQ model at one-loop. That is, assuming flavor structure in \cref{eq:yUhat} for $D_i$ leads to $\kappa_{sd}^\phi$ and $\kappa_{ds}^\phi$ that are not finite, unlike the contributions from up-quark VLQs. 
Naively, this contradicts the intuition that penguin diagrams in \cref{fig:penguin-down} match above the scale of EW symmetry breaking onto dimension five operators of the form $\phi \bar{Q}_L H d_R$, and thus should be irrelevant in the IR. This UV dependence is due to the fact that the one-loop diagram is not the 1PI and can be decomposed, by cutting the VLQ line, into two sub-diagrams, see \cref{fig:UVpenguin} (left). As a result, the counter-terms of the $\bar{D}_L d_R \phi$ and $\bar{Q}_LHD_R$ (marginal) operators contribute to $\phi\bar{Q}_L H d_R$. This is qualitatively different from the SM Higgs flavor-changing-neutral current (FCNC) which originates from the dimension-six $H^\dagger H\bar{Q}_L H d_R$ operator that is induced by the 1PI diagrams, as shown in \cref{fig:UVpenguin} (right), and thus does not require renormalization~\cite{Botella:1986hs}.

\FloatBarrier 

%%%%%%%%%%%%%%%%%%%%%%%%%%%%%%%%%%%%
%\bibliographystyle{JHEP}
\bibliographystyle{utphys28mod}
\bibliography{K2piphi}
%%%%%%%%%%%%%%%%%%%%%%%%%%%%%%%%%%%%

\end{document}